\documentclass[aps,pre,amsmath,amssymb,reprint,groupedaddress,superscriptaddress]{revtex4-1}

\usepackage{graphicx,color}% Include figure files
\usepackage{dcolumn}% Align table columns on decimal point
\usepackage{bm}% bold math
\begin{document}

\title{Laboratory experiment and discrete-element-method simulation \\
of granular-heap flows under vertical vibration}

\author{Daisuke Tsuji}
\affiliation{Department of Earth and Environmental Sciences, Nagoya University, Furocho, Chikusa, Nagoya 464-8601, Japan}

\author{Michio Otsuki}
\affiliation{Graduate School of Engineering Science, Osaka University, 1-3 Machikaneyama, Toyonaka, Osaka 560-8531, Japan}

\author{Hiroaki Katsuragi}
\affiliation{Department of Earth and Environmental Sciences, Nagoya University, Furocho, Chikusa, Nagoya 464-8601, Japan}

\date{\today}

\begin{abstract}
Granular flow dynamics on a vertically vibrated pile is studied by means of both laboratory experiments and numerical simulations.
As already revealed, the depth-averaged velocity of a fully-fluidized granular pile under strong vibration, which is measured by a high-speed laser profiler in the experiment, can be explained by the nonlinear diffusion transport model proposed by our previous paper (Tsuji ${\it et~al.}$, ${\it Phys.~Rev.~Lett.}$~${\bm 120}$, 128001 (2018)).
In this paper, we report that a similar transport model can be applied to the relation between the surface velocity and slope in the experiment.
These facts are also reproduced by particle-scale numerical simulations based on the discrete element method.
In addition, using these numerical results, the velocity profile inside the fluidized pile is measured.
As a result, we show that the flow velocity decreases exponentially with depth from the surface of the pile, which means that a clearly fluidized region, also known as shear band structure, is localized around the surface.
However, its thickness grows proportionally with the local height of the pile, i.e., the shear band does not consist of a fluidized layer with a constant thickness.
From these features, we finally demonstrate that the integration of this exponentially-decreasing velocity profile is consistent with the depth-averaged velocity predicted by the nonlinear diffusion transport model.
\end{abstract}

\maketitle

\section{Introduction}
\label{sec:Introduction}
Granular avalanches present peculiar liquid-like behavior, which has led to active discussions so far.
A dense surface flow, which is also called a heap flow, is one of the avalanche types that we frequently encounter in daily life.
The steady state of heap flows can be observed mainly in two situations: when granular media are continuously supplied onto the top of a static pile \cite{Lemieux2000,Komatsu2001, GDR2004,Jop2005, Jop2006, Katsuragi2010} or filled in a rotating drum \cite{GDR2004, Gray2001, Bonamy2002, Pont2005, Yang2008, Kleinhans2011, Amon2013, Swisher2014}.
In most of the experimental observations, the typical fluidized layer consists of $10^0-10^1$ grain diameters, which is spontaneously determined by the system itself.
The profile of the flow velocity shows an exponential decay with depth from the surface layer~\cite{Lemieux2000,Pont2005,Katsuragi2010}.
Heap flows are not in general produced unless slopes exceed more or less the angle of repose $\theta_{\rm c}$.
This is however not the case when a system is subjected to perturbation.

Mechanical vibration is, above all, frequently used as a way to cause failure of frictional contacts among grains, which changes the rheological properties of granular media in a dramatic way.
In a vibrating system, we can observe various phenomena peculiar to granular matter, such as convection \cite{Yamada2014}, segregation \cite{Breu2003}, compaction \cite{Iikawa2015}, and friction weakening \cite{Caballero-Robledo2009}.
Of course, heap flows are also induced by vibration.
It is a well known fact that granular heap structure shows relaxation to a horizontally-flat surface when subjected to strong vibration, even if the slope is clearly less than $\theta_{\rm c}$ \cite{Jaeger1989, Sanchez2007, Swisher2014, Khefif2018, Gaudel2018}.
This type of relaxation is ubiquitous in many natural phenomena.
For instance, it has been claimed in planetary science that the granular-heap relaxation governs the terrain development on astronomical objects covered with granular beds called regolith (e.g., asteroid Itokawa)~\cite{Richardson2005,Michel2009}.
Since the stability of a granular bed is dominated by a competition between gravity and vibration in many cases, a dimensionless parameter $\Gamma=A(2\pi f)^2/g$ \cite{Evesque1989} is often used to characterize the occurrence condition of the above phenomena, where $A$ and $f$ are the amplitude and frequency of imposed vibration and $g$ is gravitational acceleration.

Our previous study~\cite{Tsuji2018} has proposed the nonlinear diffusion transport (NDT) model in order to describe the behavior of granular particles driven by vertical vibration.
The NDT model is derived on a basis of simple laboratory experiments and the energy equipartition model proposed by Roering {\it et al.}~\cite{Roering1999}.
In Ref.~\cite{Tsuji2018}, we agitated a granular pile with relatively strong vertical vibration ($\Gamma\geq2$) and measured the heap flow property under a vibrating system.
By assuming the uniform fluidization of the entire pile~(cf. \cite{Sanchez2007}), the NDT model for the depth-averaged velocity of a vibro-fluidized granular bed $\bar{v}_{\rm t}$ is given by
\begin{equation}
\bar{v}_{\rm t} = \frac{cv_{\rm{vib}}}{\mu^2}\frac{|\nabla h|}{1-(|\nabla h|/\mu)^2},
\label{eq:NDT_model}
\end{equation}
where $|\nabla h|$, $\mu$, and $v_{\rm{vib}}$ are the slope of a pile, bulk friction coefficient, and maximum vibration velocity $A(2\pi f)$.
This model also introduces a parameter $c$, which indicates the conversion efficiency from vertical vibration energy into horizontal granular transport energy.
Furthermore, it turned out that this conversion efficiency $c$ is roughly constant independent of any experimental condition, implying the existence of universality.

Although the bulk flow property can be described from a macroscopic point of view in Ref.~\cite{Tsuji2018}, the NDT model cannot predict particle-scale behavior, such as velocity profile inside a fluidized pile.
This limitation is also closely related to the following questions: 
Why is the depth-averaged velocity (Eq.~(\ref{eq:NDT_model})) determined by only the slope $|\nabla h|$ when $\mu$ and $v_{\rm vib}$ are fixed?
On the analogy of conventional granular flows down an inclined plate (e.g., \cite{Pouliquen1999,Silbert2001,Andreotti2013, Gray2014, Gaudel2016}), it might be more natural that the depth-averaged velocity also depends on the height of the pile.
Besides, is the condition assumed in the derivation of the NDT model true that the entire granular bed is uniformly fluidized under vertical vibration?
In a usual gravity-driven flow on a pile, actively flowing region is limited in the vicinity of the surface~\cite{Lemieux2000,Komatsu2001,Katsuragi2010,Pont2005}.
There are also some experimental reports that a creep (very slowly moving) region exists even in a dense granular system subjected to vibration with $\Gamma\geq2$ (e.g., \cite{Yamada2014}). 

This study aims to develop a better understanding of heap flow dynamics by the mutually complementary analysis of laboratory experiments and particle-scale simulations.
In the former analysis, after reviewing the work of Ref.~\cite{Tsuji2018} briefly, the surface flow property is analyzed by the pattern matching of profiles using the same data, and then compared to the depth-averaged flow property.
In the latter analysis, numerical simulations are utilized to obtain the information that cannot be accessed from experimental data.
Particularly, the velocity profile inside the pile is extensively studied from a microscopic point of view in contrast to Ref.~\cite{Tsuji2018}.
Finally, in order to prove the consistency between the NDT model (Eq.~(\ref{eq:NDT_model})) and particle-scale dynamics investigated in this paper, the depth-averaged velocity is deduced by integrating the internal velocity profile.

\section{Setup}
\label{Setup}

This section firstly introduces the experimental configuration and the process of data acquisition. Then, we explain the setup of the numerical simulation, which attempts to reproduce the experiment, and the analysis method of those numerical data. 

\subsection{Laboratory experiment}
\label{sec:Laboratory experiment}

\begin{figure*}[htbp]
\begin{center}
\includegraphics[width=6.5 in]{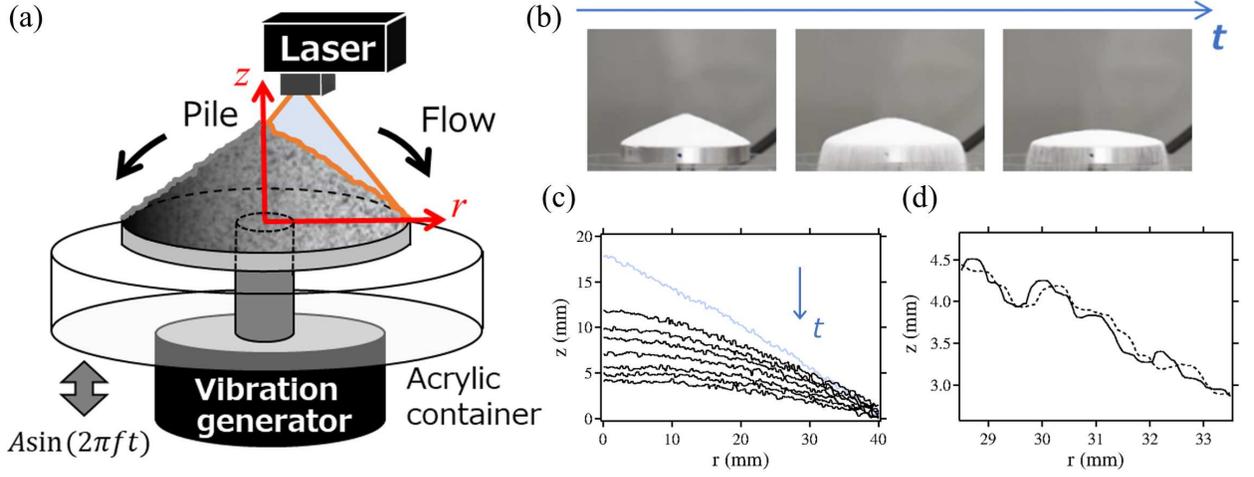}
\end{center}
\caption{(a) Schematic illustration of the experimental system.
(b) Snapshots of a relaxing pile during vibration.
(c) Raw data of radial profiles taken by a laser profiler. The experimental conditions are $A = 0.04~{\rm mm}$ and $f = 200~{\rm Hz}$ $(\Gamma = 6)$. A light blue curve shows a profile of an initial pile with the angle of repose, which is taken before turning on a vibration generator, corresponding to $t=-T_{\rm a}$. Black profiles are taken at $t=0, 0.26, 0.5, 1.0, 2.0, 3.0, 5.0$~s from top to bottom. Note that the data between light blue and top black profiles are not used in this study as the vibration amplitude $A$ is not constant.
(d) Example of a magnified view of profiles. Solid and dashed profiles correspond to $t=0.78~{\rm and}~0.8$~s. It seems that grains flow nearly holding the profile pattern.
}
\label{fig:exp_system}
\end{figure*}

The experimental method and data used in this study are the same as Ref.~\cite{Tsuji2018}.
A schematic illustration of the experimental system is shown in Fig.~\ref{fig:exp_system}(a).
A conical granular pile with the angle of repose is created on a disk with radius $R=40~{\rm mm}$, which is horizontally mounted onto an electromechanical vibrator (EMIC 513-B/A).
The granular materials used in the experiment are listed in Table \ref{tab:grain_property}.
The surface of the disk is pasted with the same type of grains so that the heap structure can be sustained.
After the creation of a pile, sinusoidal vertical vibration is continuously applied to the disk.
Here, the amplitude is gradually increased during the initial short term $T_{\rm a}=0.5~{\rm s}$ to calmly reach stable vibration conditions without burst signals. 
The time when the vibration achieves a stable state is defined as $t = 0~{\rm s}$. 

The amplitude $A$ and frequency $f$ are varied in the range of $10^{-4}-10^{0}~{\rm mm}$ and $50-500~{\rm Hz}$ such that granular layers are fluidized. 
Actually, the onset criterion of fluidization occurrence is difficult to determine. 
Although $\Gamma$ seems to be the most relevant, the critical condition $\Gamma_{\rm c}$ fluctuates around $1 < \Gamma_{\rm c} < 2$ depending on $f$~\cite{Tennakoon1998,King2000}.
To avoid such complexity and focus only on clearly fluidized regimes, in this study a pile is subjected to relatively strong vibration of $\Gamma=2 - 10$.

Once granular media begin to flow to the outside of the disk, the shape of the pile gets relaxed (Fig.~\ref{fig:exp_system}(b)). 
Outflowed grains are captured by an acrylic container surrounding the disk. 
The advantage of this experimental configuration is that the sidewall effect, which must be taken into consideration in the case of quasi-2D flows \cite{Jop2005,Jop2006}, does not appear at all.

In order to measure the flow properties during the vibration, surface profiles of the relaxing pile are recorded by a high-speed laser profiler (KEYENCE LJ-V7080). 
Figure~\ref{fig:exp_system}(c) shows profiles taken in the experimental condition of $A = 0.04~{\rm mm}$ and $f = 200~{\rm Hz}$ $(\Gamma = 6)$ with Material 1 in Table \ref{tab:grain_property}.
The sampling rate is 50 Hz, and the horizontal spacial resolution is 50 $\mu$m/pix. 
The number of pixels is $800$, which can just cover all of profiles at $r = 0-40 {\rm~mm}$, where $r$ is the radial distance from the center of the disk.
The origin of the height coordinate $z=0$ is calibrated to the surface of the disk.
The measurement error of the laser along the $z$ direction is less than $50~\mu {\rm m}$.
These high resolution and accuracy of the laser measurement enable us to precisely capture the grain-scale movement as shown in the magnified view of profiles (Fig.~\ref{fig:exp_system}(d)), where we can see that the surface layer moves almost keeping its profile pattern.
Experiments were performed three times for each set of conditions in order to check the reproducibility.
Henceforth, the experimental conditions shown in Figs.~\ref{fig:exp_system}(b)$-$(d) are used for subsequent plots unless otherwise noted.

\begin{table}[htbp]
\setlength{\tabcolsep}{2mm}
\begin{center}
\begin{tabular}{lcccc}
\hline
Material   	  &  $d$~(mm)     &	$\rho$~(g/cm$^3$)	& tan$\theta_{\rm c}$ 	& Note	 \\
\hline
1. Alumina ball    &	$0.5\pm0.1$ &	3.9		&	$0.45$	   & A.O.	 \\	%AS-ONE AL9-0.5
2. Alumina ball    & 	$1.0\pm0.1$ &	3.9		&	$0.45$ 	& A.O.  \\	%AS-ONE AL9-1
3. Zirconia ball  &	$0.5\pm0.1$ &	5.9		&	$0.46$    & A.O.	 \\	%AS-ONE CZC0050
4. Rough sand	  &	$1.0\pm0.3$ &	2.6		&	$0.65$    & JIS	 \\	%JIS R 5201
\hline
\end{tabular}
\caption{Granular media used in the experiments. A.O. and JIS represent AS ONE (supplier of materials) and Japanese Industrial Standards.}
\label{tab:grain_property}
\end{center}
\end{table}

\subsection{Numerical simulation}
\label{sec:Numerical simulation}

\subsubsection{Contact model}
\label{sec:Contact model}

Particle-scale simulations are conducted in both two dimensions (2D) and three dimensions (3D) by means of the discrete element method (DEM) \citep{Cundall1979}.
Modeled in this study are $N$ polydisperse disks/spheres of constant material density with the maximum diameter $d$ and mass $m$.
A small polydispersity equally ranging from $d/\sqrt{2}$ to $d$ is given in order to prevent the crystallization of a system~\citep{Iikawa2016}.
The position, velocity, and angular velocity of $i$th particle are denoted by $\bm{r}_i$, $\bm{v}_i$, and $\bm{\omega}_i$, respectively.
The total force applied on $i$th particle is determined by a combination of gravity and the contact force with $j$th particle $\bm{f}_{ij}$, which consists of the normal part $\bm{f}_{ij}^{\rm (n)}$ and tangential part $\bm{f}_{ij}^{\rm(t)}$, i.e., $\bm{f}_{ij}=\bm{f}_{ij}^{\rm (n)}+\bm{f}_{ij}^{\rm (t)}$.

For a pair of two contacting $i$th and $j$th particles with diameters $d_i$ and $d_j$, the normal compression $\delta_{ij}$ and relative velocity ${v}_{ij}^{\rm (n)}$ are given by 
\begin{equation}
\delta_{ij}=\left(d_{i} + d_{j}\right)/2 - \left|\bm{r}_{ij}\right|,
\label{eq:def_normal_compression}
\end{equation}
\begin{equation}
{v}_{ij}^{\rm (n)}=\bm{v}_{ij}\cdot\bm{n}_{ij},
\label{eq:def_normal_relative_velocity}
\end{equation}
where $\bm{r}_{ij}=\bm{r}_{i}-\bm{r}_{j}$, $\bm{v}_{ij}=\bm{v}_{i}-\bm{v}_{j}$, $\bm{n}_{ij}=\bm{r}_{ij}/\left|\bm{r}_{ij}\right|$.
The normal contact force $\bm{f}_{ij}^{\rm (n)}$ is modeled using a normal spring constant $k^{\rm (n)}$ and viscous constant $\eta^{\rm (n)}$ as
\begin{equation}
\bm{f}_{ij}^{\rm (n)} = \Theta\left(\delta_{ij}\right)\left(k^{\rm (n)}\delta_{ij}-\eta^{\rm (n)}{v}_{ij}^{\rm (n)}\right)\bm{n}_{ij},
\label{eq:normal_force}
\end{equation}
where $\Theta(x)$ is the Heaviside step function, i.e., $\Theta(x)=1$ for $x\geqslant0$ and $\Theta(x)=0$ otherwise.

On the other hand, the tangential contact force $\bm{f}_{ij}^{\rm (t)}$ is modeled with a tangential spring constant $k^{\rm (t)}$ as 
\begin{equation}
\bm{f}_{ij}^{\rm (t)}=-k^{\rm (t)}\bm{u}_{ij},
\label{eq:def_tangential_force}
\end{equation}
where the tangential displacement $\bm{u}_{ij}$ is obtained by integrating the tangential relative velocities $\bm{v}_{ij}^{\rm (t)}$, which are expressed as
\begin{equation}
\bm{u}_{ij}=\int_{\rm stick} \left( \bm{v}_{ij}^{\rm (t)} - \frac{(\bm{u}_{ij} \cdot \bm{v}_{ij})\bm{r}_{ij}}{\left| \bm{r}_{ij} \right|^2} \right) dt,
\label{eq:def_tangential_compression}
\end{equation}
\begin{equation}
\bm{v}_{ij}^{\rm (t)}=\left(\bm{v}_{ij}\cdot\bm{t}_{ij}\right)\bm{t}_{ij} -\frac{1}{2}\left(d_{i}\bm{\omega}_{i}+d_{j}\bm{\omega}_{j}\right)\times\bm{n}_{ij},
\label{eq:def_tangential_relative_velocity}
\end{equation}
where $\bm{t}_{ij}=\left(-r_{ij,y}/\left|\bm{r}_{ij}\right|,r_{ij,x}/\left|\bm{r}_{ij}\right|\right)$, and the second term in the integral of Eq.~(\ref{eq:def_tangential_compression}) insures that $\bm{u}_{ij}$ always lies on the tangent plane of the contact point. 
In Eq.~(\ref{eq:def_tangential_compression}), ``stick'' means that the integral is performed only when $\left|\bm{f}_{ij}^{\rm(t)}\right|<\mu_{m}\left|\bm{f}_{ij}^{\rm (n)}\right|$ is satisfied, where $\mu_m$ is a microscopic friction coefficient.
This condition indicates that the Coulomb friction criterion holds in quasistatic motion:
$\left|\bm{f}_{ij}^{\rm(t)}\right|=k^{\rm (t)}|\bm{u}_{ij}|$ in the ``stick'' region of $\left|\bm{u}_{ij}\right| < \mu_{m}\left|\bm{f}_{ij}^{\rm(n)}\right|/k^{\rm (t)}$, while $\left|\bm{f}_{ij}^{\rm(t)}\right|$ remains $\mu_{m}\left|\bm{f}_{ij}^{\rm(n)}\right|$ in the ``slip'' region of $\left|\bm{u}_{ij}\right| \geq \mu_{m}\left|\bm{f}_{ij}^{\rm(n)}\right|/k^{\rm (t)}$.
In addition, the torque $\bm{T_i}$ of $i$th particle is given by
\begin{equation}
\bm{T}_i=-\sum_j \frac{d_i}{2}\bm{n}_{ij}\times\bm{f}_{ij}^{\rm(t)}.
\label{eq:def_torque}
\end{equation}

Using the contact forces introduced above, the translational and rotational accelerations of $i$th particle are determined by Newton's second law:
\begin{equation}
m_{i}\ddot{\bm{r}_{i}}=\sum_j \bm{f}_{ij}+m_{i}\bm{g},
\label{eq:newton_translation}
\end{equation}
\begin{equation}
I_{i}\dot{\bm{\omega}_{i}}=\bm{T}_i,
\label{eq:newton_rotation}
\end{equation}
where $m_{i}$ and $I_{i}$ are the mass and moment of inertia of $i$th grain, and $\bm{g}=(0,0,-g)$ is gravity.
In our simulation, $d$, $m$, and $g$ are set to be unity, and all of the quantities are computed in dimensionless forms, where the unit time is $\sqrt{d/g}$.
After the computation, we give dimensions to all of quantities using the same units as the experiment $(d=0.5~{\rm mm}, m=~2.5\times10^{-4}~{\rm g}, ~{\rm and}~g=9800~{\rm mm/s^2})$.

Another important point in DEM simulations is how to choose mechanical parameters.  
This study adopts sufficiently hard spheres/disks ($k^{\rm (n)}=10^{4}~[mg/d]$) so that this choice does not have an influence on the simulation result.
In fact, the simulation result changes mostly only within error even if $k^{\rm (n)}=10^{5}~{\rm or}~10^6~[mg/d]$ is used.
The ratio $k^{\rm (t)}/k^{\rm (n)}$ depends on the material property of particles, which is typically set to be $k^{\rm (t)}/k^{\rm (n)}=2/7-2/3$ in DEM simulations.
However, we found that the result is not sensitive to this ratio in this study.
Although all of the results shown below are obtained with $k^{\rm (t)}/k^{\rm (n)}=2/7$, the data using $k^{\rm (t)}/k^{\rm (n)}=2/5~{\rm or}~2/3$ change only within error as well.
In contrast, a microscopic friction coefficient $\mu_m$ has a little influence on the result.
In this study, relatively large friction $\mu_{m}=0.8$ is used, the reason of which is explained in Sec.~\ref{sec:Simulation procedure}.

The energy loss due to inelastic collisions is characterized by the coefficient of restitution $e$, which is defined as the ratio of the post-collisional to pre-collisional normal relative velocity and also can be analytically written as
\begin{equation}
e = {\rm exp}\left(-\eta^{\rm (n)} t_{\rm col}/m \right),
\label{eq:def_restitution_coef}
\end{equation}
where $t_{\rm col}$ is the collision time:
\begin{equation}
t_{\rm col} = \frac{\pi}{\sqrt{2k^{\rm(n)}/m - {\eta^{\rm(n)}}^2/m^2 }}
\label{eq:def_collision_time}
\end{equation}
To investigate the dependence on the degree of inelasticity, we vary $e$ from $0.7$ to $0.9$ by arranging $\eta^{\rm (n)}$, which corresponds to typical granular matter.

Last but not least, care must be taken for the time step of the calculation $\delta t$.
In our simulation, the bulk flow property is not sensitive to $\delta t$ once it becomes less than $t_{\rm col}/20$. 
Therefore, all of the simulations in this study are conducted with the time step $\delta t < t_{\rm col}/20$.

\subsubsection{Simulation procedure}
\label{sec:Simulation procedure}

Here, the simulation procedure from the creation of a pile to vibration is explained.
Firstly, particles are randomly filled into a triangle (2D) or cone (3D) space on a fixed base.
The base is composed of the same grains with diameter $d$, which are placed without gaps, and the system size is, unless otherwise noted, the same as the experiment ($R=40$~mm). 
The initial packing fraction $\phi_{\rm init}$ is set to be slightly smaller than the jamming point~\cite{Otsuki2009}: $\phi_{\rm init}=0.75~{\rm (2D)}~{\rm and}~0.55~{\rm (3D)}$.
The slope of a filled triangle/cone is always larger than the angle of repose $\theta_{\rm c}$. 
The number of simulated particles is $\sim 5\times10^3~{\rm (2D)}~{\rm and}~\sim 5\times10^5~{\rm (3D)}$.

Next, by imposing gravity to all of particles simultaneously, a pile with the angle of repose is spontaneously created.
Since simulated particles are completely spherical, the angle of repose $\theta_{\rm c}$ produced in DEM simulations is smaller than that of real grains.
In addition, $\theta_{\rm c}$ slightly depends on a microscopic friction coefficient $\mu_{m}$~\cite{Zhou2001, Zhou2002}.
In this study, $\theta_{\rm c}$ increases with $\mu_m$, and almost saturates in the range of $\mu_{m} \geq 0.8$.
We have confirmed that the flow property does not change once $\mu_{m}$ exceeds $0.8$ as the bulk frictional property does not change either.
In order to reproduce as realistic a pile as possible, $\mu_m=0.8$ is employed in this study, where ${\tan \theta_{\rm c}}=0.35~({\rm 2D})~{\rm and}~0.40~({\rm 3D})$.

\begin{figure*}[htbp]
\begin{center}
\includegraphics[width=6.8 in]{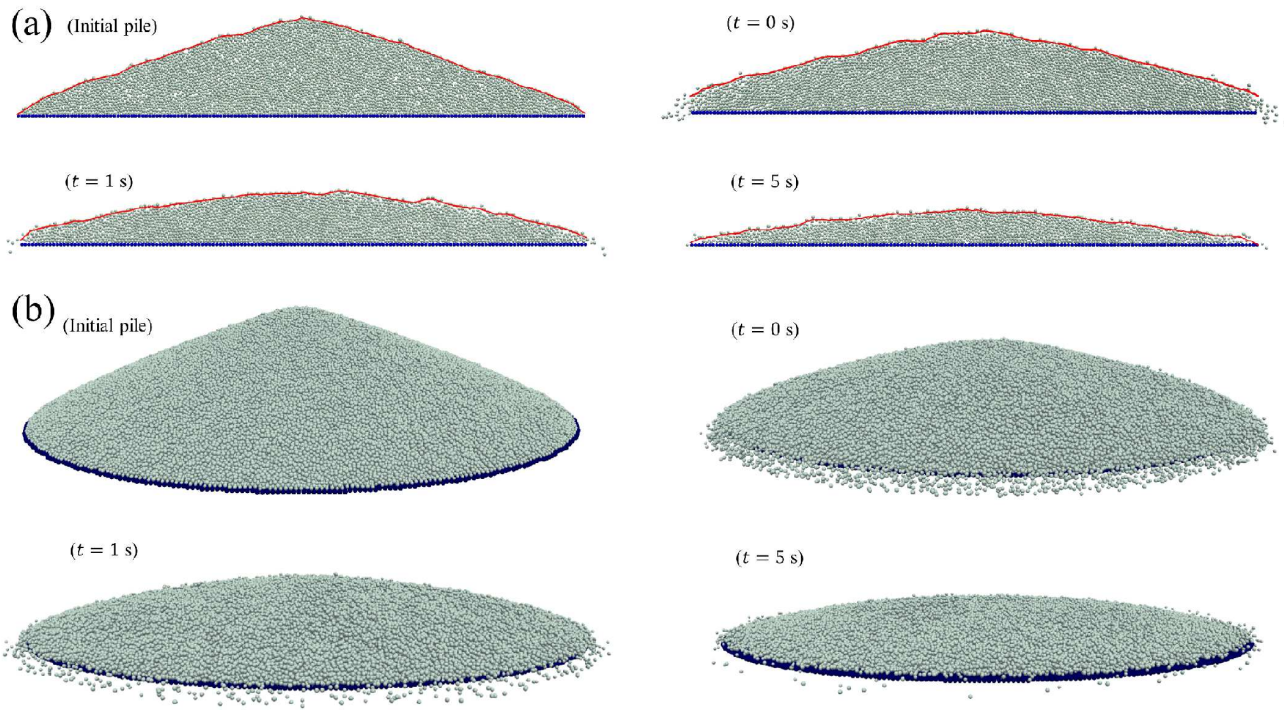}
\end{center}
\caption{Reproduction of the experiment by the DEM simulation.
Panels (a) and (b) depict the 2D and 3D simulations, respectively.
Only vibrating bottom particles are colored in dark blue.
The first data are taken before applying vibration ($t=-T_a$), and the others are taken at $t=0,1,5~{\rm s}$.
Red curves in Panel (a) are surface profiles computed by the method explained in Sec.~\ref{sec:Production of surface profiles}. 
}
\label{fig:DEMpile}
\end{figure*}

Finally, particles at the base are vertically vibrated with $A \sin (2\pi f)t$ in the same way as the experiment.
Note that we use a smaller amplification period ($T_{\rm a}=0.25$~s) than laboratory experiments so that the flow property can be measured in broad $|\nabla h|$ ranges.
The time evolution of a simulated vibrating pile is shown in Fig.~\ref{fig:DEMpile} for both 2D and 3D cases.
The vibration condition is same as Fig.~\ref{fig:exp_system} and grains with $e=0.8$ are used, which are also used for subsequent plots unless otherwise specified.
More than $10$ and $5$ simulation runs for each set of conditions were conducted for the 2D and 3D cases, respectively.

\subsubsection{Production of surface profiles}
\label{sec:Production of surface profiles}

From the obtained DEM data, the surface profile is produced to confirm the consistency with the experiment.
The resolution along the radial direction is set to be $1d$, in which the position of the highest-located particle is given as the actual height $h_{\rm act}$.
However, this raw profile scatters due to the saltation motion of particles, which could lead to a large error on the result.
Therefore, to stabilize the data, we take the moving-average with a $5d$ window in the following way:
\begin{equation}
h(r)=\frac{1}{5}\sum_{n=r/d-2}^{r/d+2}h_{\rm act}(nd).
\label{eq:moving_average}
\end{equation}
As an example, the surface profiles produced by Eq.~(\ref{eq:moving_average}) are drawn onto 2D piles in Fig.~\ref{fig:DEMpile}(a), which are in good agreement with surface particles.

\section{Experimental result}
\label{sec:Experimental result}

\begin{figure*}[htbp]
\begin{center}
\includegraphics[width=6.5 in]{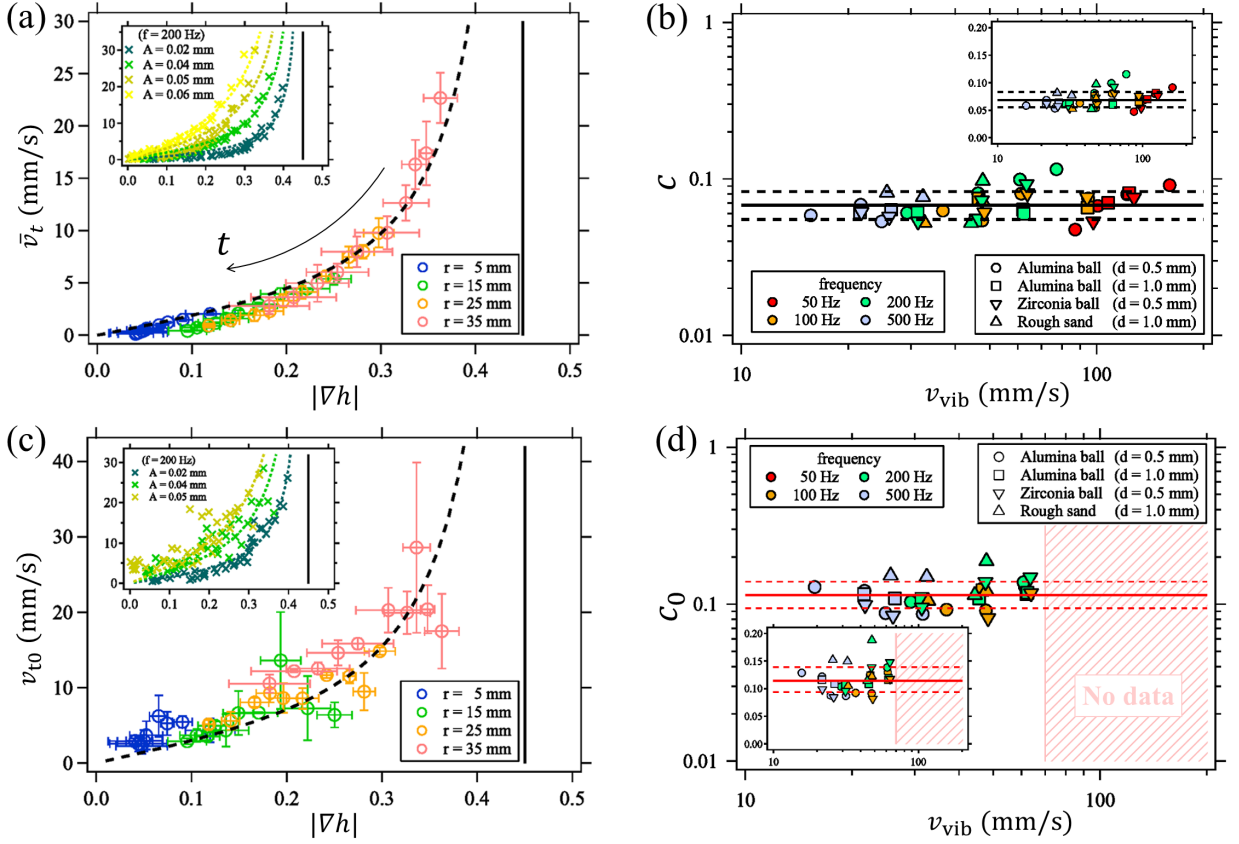}
\end{center}
\caption{Laboratory experiment result:~
(a) Depth-averaged velocity $\bar{v}_{\rm t}$ versus slope $|\nabla h|$.
The data points shift to a smaller $|\nabla h|$ range as time passes.
Colors represent analysis positions.
The vertical solid line corresponds to $|\nabla h|=\tan \theta_c$, where $\bar{v}_{\rm t}$ diverges.
A dashed curve shows the best fitting by Eq.~(\ref{eq:NDT_model}).
Inset: Analysis results for various vibration strength.
The axes are the same as the main panel.
All of the data are also fitted by Eq.~(\ref{eq:NDT_model}).
(b) Conversion efficiency $c$ as a function of the maximum vibration velocity $v_{\rm vib}$. 
Colors and symbols represent vibration frequency $f$ and materials used (Table.~\ref{tab:grain_property}). 
Solid and dashed lines show $c=0.068$ with $1\sigma =  0.014$.
The inset shows the same data in linear scale.
(c) Surface velocity $v_{\rm t0}$ versus slope $|\nabla h|$.
A dashed curve and solid line show the best fitting by Eq.~(\ref{eq:NDT_model_surface}) and the divergent point $|\nabla h|=\tan \theta_c$, respectively.
Inset: Analysis results for various vibration strength.
The axes are the same as the main panel.
All of the data are also fitted by Eq.~(\ref{eq:NDT_model_surface}).
(d) Fitting parameter of Eq.~(\ref{eq:NDT_model_surface}) $c_{\rm 0}$ as a function of the maximum vibration velocity $v_{\rm vib}$. 
The value of $c_{\rm 0}$ physically corresponds to the conversion efficiency of the surface flow.
Red solid and dashed lines draw $c_{\rm 0}=0.11$ with $1\sigma =  0.022$.
In the range of $v_{\rm vib}>70~{\rm mm/s}$, which is the area hatched in light red,  the pattern matching algorithm cannot be used due to the strong fluctuation of profiles.
The inset shows the same data in linear scale.
}
\label{fig:exp_result}
\end{figure*}

\subsection{Depth-averaged velocity}
\label{sec:Depth-averaged velocity}

First, the consistency between the NDT model (Eq.~(\ref{eq:NDT_model})) and experimental data, which has been confirmed by Ref.~\cite{Tsuji2018}, is briefly reviewed.
According to the NDT model, the depth-averaged velocity along the horizontal direction $\bar{v}_{\rm t}$, defined as 
\begin{equation}
\bar{v}_{\rm t}=\frac{1}{h}\int_0^h v_{\rm t}(z)dz,
\label{eq:depth_averaged0}
\end{equation}
where $v_{\rm t}(z)$ is flow velocity at position $z$, is a function of only slope $|\nabla h|$ when the other experimental conditions such as $\mu$ and $v_{\rm vib}$ are fixed.
To check this, $\bar{v}_{\rm t}$ and $|\nabla h|$ are measured at four different points $(r=5,15,25,35~{\rm mm})$ and various time ($t=0.1\times \sqrt{2}^n~{\rm s}~(n=0,1,2,\cdots)$).
$\bar{v}_{\rm t}$ can be computed by the volume flux $q$ divided by the height $h$ for each position $r$ and time $t$, i.e., $\bar{v}_{\rm t}(r,t)=q(r,t)/h(r,t)$, where $q(r,t)$ is defined as the volume of granular media flowing across a unit length per unit time and can be calculated from the volume change of the relaxing pile as
\begin{equation}
q(r,t)=\frac{1}{r dt} \int_0^{r} |h(r', t+dt)-h(r', t)|r' dr'.
\label{eq:flux_definition}
\end{equation}
$|\nabla h|$ is locally measured by the linear least-squares method using the neighbor profiles of $r\pm 5~{\rm mm}$.
The detailed measurement method of these quantities is explained in Ref.~\cite{Tsuji2018} and its supplementary material.

In Fig.~\ref{fig:exp_result}(a), $\bar{v}_{\rm t}$ is plotted against $|\nabla h|$, where all of the data are collapsed into a single curve and can be fitted by Eq.~(\ref{eq:NDT_model}).
As shown in the inset of Fig.~\ref{fig:exp_result}(a), this scaling can also be observed for experimental data obtained in different experimental conditions. 
Since Ref.~\cite{Tsuji2018} assumes $\mu=\tan \theta_{\rm c}$ for the sake of simplicity, the fitting parameter is only $c$ here.
The dependence of $c$ on various experimental conditions is also shown in Fig.~\ref{fig:exp_result}(b).
Interestingly, $c$ is almost not sensitive to any experimental condition, such as the vibration condition and type of granular material.
This means that the conversion efficiency from vertical vibration energy into horizontal granular transport energy could be a universal constant.

\subsection{Surface velocity}
\label{sec:Surface velocity}
Next, we investigate whether the NDT model holds in particle scale.
By the pattern matching of subsequent profile images, the velocity of a surface flow is measured at the same spatial and temporal resolutions as the depth-averaged velocity (Fig.~\ref{fig:exp_result}(a)).
The applied algorithm is similar to the particle image velocimetry (PIV) method~\cite{Lueptow2000}, although the profile data of this study is one dimensional.
Small spacial windows for the profile pattern matching are given by $\pm5~{\rm mm}$ at each position $(r=5,15,25,35~{\rm mm})$ as well as computation of slope $|\nabla h|$. 
Then, adequate and appropriate time interval $\Delta t$ is chosen so that the distance of displacement can be clearly identified as can be seen in Fig.~\ref{fig:exp_system}(d).
This distance is determined by finding the position where one-dimensional cross-correlation function between two profiles shows the maximum value.
Here, we define $\Delta r$ as the $r$ component of this displacement (not along the surface) as the depth-averaged velocity $\bar{v}_{\rm t}$ is measured along the $r$ direction as well.
Consequently, the surface velocity along the $r$ direction $v_{\rm t0}$ can be estimated by $\Delta r/\Delta t$.
The range of $\Delta t$ is properly chosen in order for the typical $\Delta r$ to be a few grain diameter.
The detail of this pattern matching is summarized in Appendix~\ref{appendixA}.

Figure~\ref{fig:exp_result}(c) shows the comparison between $v_{\rm t0}$ and $|\nabla h|$, where the data are scaled by a single curve as well as $\bar{v}_{\rm t}$ in Fig.~\ref{fig:exp_result}(a).
The data can also be fitted by the NDT model:
\begin{equation}
v_{\rm t0} = \frac{c_{\rm 0}v_{\rm{vib}}}{\mu^2}\frac{|\nabla h|}{1-(|\nabla h|/\mu)^2},
\label{eq:NDT_model_surface}
\end{equation}
where $c_{\rm 0}$ is a constant, physically corresponding to the conversion efficiency of the surface flow.
The dependence of $c_{\rm 0}$ on experimental conditions is shown in Fig.~\ref{fig:exp_result}(d).
Note that the data in the range of $v_{\rm vib}>70~{\rm mm/s}$ is not plotted because the variation of surface profiles is too large to apply the pattern matching algorithm in these strong vibration conditions.
Although the data range is slightly limited, the same flat trend as Fig.~\ref{fig:exp_result}(b) is observed.
However, the value of $c_{\rm 0}=0.11$ is different from $c=0.068$: 
\begin{equation}
c_{\rm 0} \approx 2c,
\label{eq:c_csurf}
\end{equation}
which suggests that the flow velocity on the top of the relaxing pile is approximately twice as large as the depth-averaged velocity; and the flow has a structure that the velocity decreases as going deeper from the surface.

This tendency is qualitatively consistent with shear band structure of conventional heap flows \cite{Katsuragi2010}, i.e., a clearly fluidized region is localized around the surface.
However, the derivation of the NDT model assumes a uniformly-fluidized granular pile \cite{Tsuji2018}.
To verify the consistency between the observed results and the NDT model, the internal velocity profile of granular flows has to be investigated.
To solve this matter, numerical simulations are much more convenient than experimental approaches for the setup of this study (Fig.~\ref{fig:exp_system}(a)).
In the next session, we will go into the discussion on numerical results.

\section{Simulation result}
\label{sec:Simulation result}

\subsection{Depth-averaged velocity}
\label{sec:Depth_averaged_velocity}

\begin{figure*}[htbp]
\begin{center}
\includegraphics[width=7.0 in]{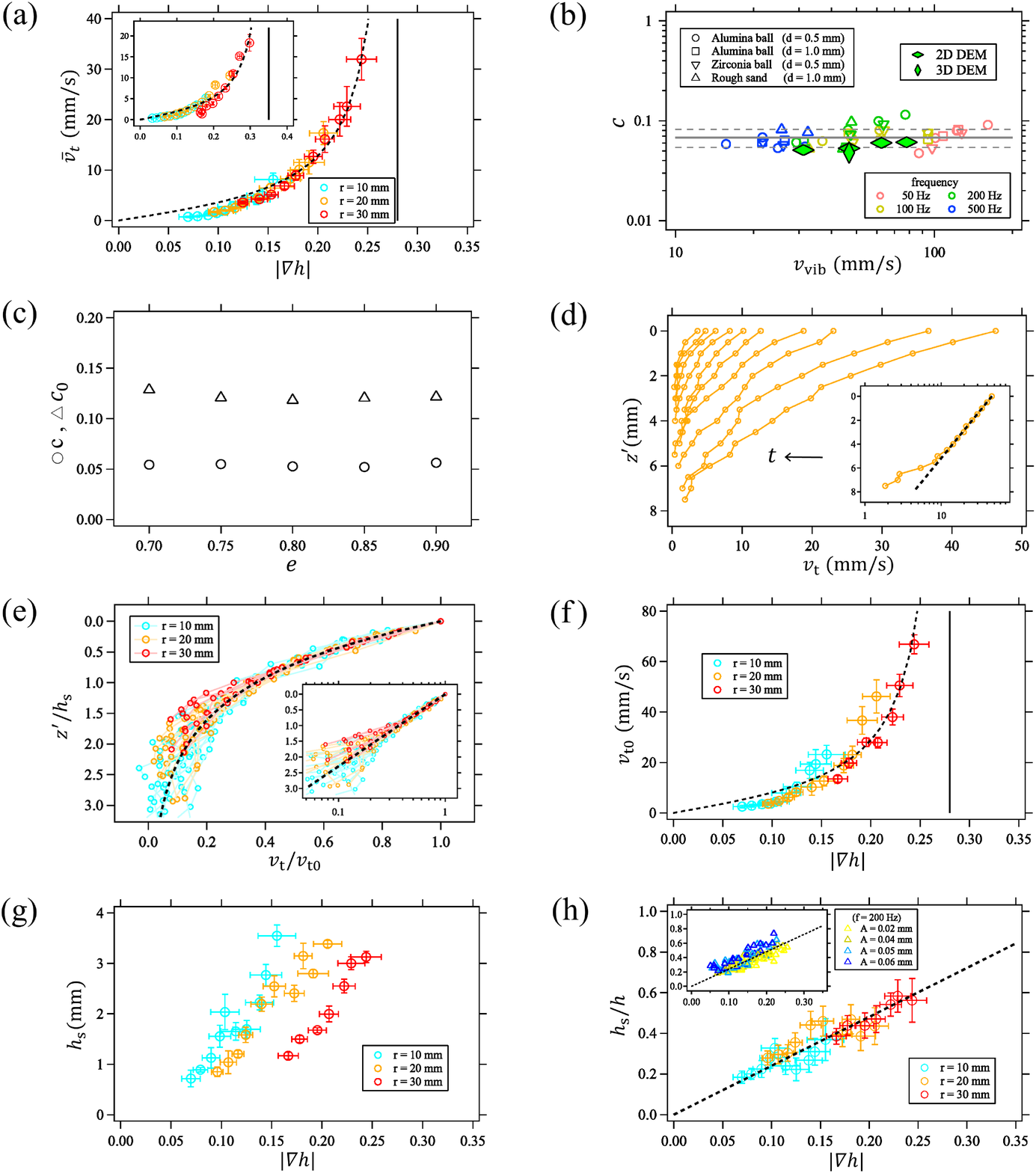}
\end{center}
\caption{Numerical simulation result:~
(a) Depth-averaged velocity $\bar{v}_{\rm t}$ versus slope $|\nabla h|$.
The main plot and inset correspond to the 2D and 3D cases, respectively.
The axes of the inset are the same as the main plot.
Both data can be fitted by Eq.~(\ref{eq:NDT_model}), which are drawn in dashed curves.
The values of $\mu$, which are determined by these fittings and drawn in vertical solid lines, are $0.28$ (2D) and $0.35$ (3D).
(b) Conversion efficiency for a bulk flow $c$ as a function of the maximum vibration velocity $v_{\rm vib}$. 
The $c$ values computed by the DEM simulations (filled diamonds) are plotted onto experimental results (open circles) for comparison.
(c) Conversion efficiencies $c$ ($\bigcirc$) and $c_{\rm 0}$ ($\bigtriangleup$) versus the coefficient of restitution $e$.
The relation $c_{\rm 0} \approx 2c$, which was empirically obtained from experimental data (Eq.~(\ref{eq:c_csurf})), is also almost satisfied independent of $e$.
(d) Velocity profiles obtained at $r=20$~mm and various time.
The velocity $v_{\rm t}$ decreases exponentially with depth $z'$.
Inset: Example of the fitting by Eq.~(\ref{eq:velocity_profile}) in semi-log plot.
The data is taken at $t=0$ s and $r=20$ mm. 
(e) All of the scaled velocity profile data, which are collapsed into a single curve: $v_{\rm t}/v_{\rm t0}=\exp(-z'/h_{\rm s})$.
The inset shows the same data in semi-log plot.
(f) Surface velocity $v_{\rm t0}$ versus slope $|\nabla h|$.
The data are fitted by Eq.~(\ref{eq:NDT_model_surface}).
(g) Characteristic thickness of a fluidized layer $h_{\rm s}$ as a function of $|\nabla h|$.
(h) Scaled characteristic thickness of a fluidized layer $h_{\rm s}/h$ as a function of $|\nabla h|$.
Inset: Analysis data for various vibration strength.
Dash lines show a linear relationship (Eq.~(\ref{eq:h0_h})).
}
\label{fig:sim_result}
\end{figure*}

The consistency with the experimental result is firstly confirmed for simulation data.
Figure~\ref{fig:sim_result}(a) shows the relation between the depth-averaged velocity $\bar{v}_{\rm t}$ and slope $|\nabla h|$ obtained in 2D and 3D simulations.
While $\bar{v}_{\rm t}$ and $|\nabla h|$ are measured at three different points $(r=10,20,30~{\rm mm})$, other measurement methods are the same as the experiment (Fig.~\ref{fig:exp_result}(a)).
Both data are collapsed into single curves, and can be fitted by Eq.~(\ref{eq:NDT_model}).
It should be noted, however, that both $\mu$ and $c$ are left as free fitting parameters.
Although $\mu$ was fixed by $\tan \theta_{\rm c}$ in the experiment~\cite{Tsuji2018}, $\mu$ can be in general different from $\tan \theta_{\rm c}$~\cite{Ghazavi2008} and also tends to decrease in the presence of vibration~\cite{Caballero-Robledo2009}.
In fact, $\mu=0.28~({\rm 2D})~{\rm and}~0.35~({\rm 3D})$ result in better agreements with the data of Fig.~\ref{fig:sim_result}(a) than $\mu={\tan \theta_{\rm c}}=0.35~({\rm 2D})~{\rm and}~0.40~({\rm 3D})$.
Besides, these fitting-based $\mu$ values are almost independent of simulation conditions (only within $\pm0.02$) in the range of $v_{\rm vib}=30-80$~mm/s and $e=0.7-0.9$. 
Thus, the fixed values $\mu=0.28~({\rm 2D})~{\rm and}~0.35~({\rm 3D})$ are employed for all of the analyses of DEM simulations. 

The $c$ values computed from the DEM simulation are plotted onto those obtained from experimental data in Fig.~\ref{fig:sim_result} (b).
It seems that there is no significant difference between simulations and experiments in terms of the $c$ values.
This result supports the fact that our DEM simulation can reproduce laboratory experiments well.
In addition, $c$ does not depend on spatial dimensions (Figs.~\ref{fig:sim_result}(a) and (b)).
Therefore, 2D simulations are used to investigate the parameter dependence and analyze the velocity profile in Sec.~\ref{sec:Velocity_profile}. 

The influence of the restitution coefficient $e$ is also investigated here.
Figure~\ref{fig:sim_result}(c) shows the relation between $c$ and $e$, which suggests that $c$ is not sensitive to $e$.
This can be understood in the following sense:
da Cruz {\it et al.}~\cite{Cruz2005} report that $e$ has no influence on the bulk frictional property in plane shear flows.
In fact, the value of $\mu$ determined by the fitting of Eq.~(\ref{eq:NDT_model}) is little influenced by $e$ as mentioned above.
Given that the frictional property does not depend on $e$ in our system, the forces considered in the derivation of the NDT model are not influenced~\cite{Tsuji2018}.
It is thus natural that the bulk flow property characterized by $\bar{v}_{\rm t}$ does not change depending on $e$.

\subsection{Velocity profile}
\label{sec:Velocity_profile}

The velocity profile inside the relaxing pile, which was technically challenging to address from experimental data, can be measured from the DEM simulation data. 
Here, we introduce the $z'$ coordinate defined by the depth from the surface ($z'=h-z$): $z'=0~{\rm and}~h$ correspond to the surface and bottom, respectively.
The velocity profiles are measured as a function of $z'$ at various time and $r=10,20,30$~mm as Fig.~\ref{fig:sim_result}(a) does.
As an example, profiles obtained at $r=20~{\rm mm}$ and various time are shown in Fig.~\ref{fig:sim_result}(d), where velocity profiles depend on not only $z'$ but also $t$ as $h$ and $|\nabla h|$ vary with $t$. 

On the analogy of steady heap flows~\cite{Komatsu2001, Katsuragi2010}, we found that the flow velocity $v_{\rm t}(z', h, |\nabla h|)$ decreases exponentially with the depth $z'$, implying the presence of shear band structure.
The data are fitted by the following function:
\begin{equation}
v_{\rm t}=v_{\rm t0}\exp\left( -\frac{z'}{h_{\rm s}} \right),
\label{eq:velocity_profile}
\end{equation}
where $v_{\rm t0}$ and $h_{\rm s}$ represent the surface velocity and the characteristic thickness of a clearly fluidized layer.
The dependence of $v_{\rm t}$ on $h$ and $|\nabla h|$ should be included in $v_{\rm t0}$ and $h_{\rm s}$.

An example of the fit by Eq.~(\ref{eq:velocity_profile}) is shown in the inset of Fig.~\ref{fig:sim_result}(d).
Despite being able to fit the data around the surface $(z'=0)$ in a good quality, misfit to the exponential curve comes to remarkably appear as going deeper.
This might be caused by the boundary effect as particles are trapped at the bottom, i.e., $v_{\rm t}\approx0~{\rm at}~z'=h$.
Besides, when $h$ is too small, an exponential profile cannot be distinguished from another form of function such as a linear profile. 
To avoid these matters, only the upper halves ($z'>h/2$) of sufficiently thick profile data ($h>5d$) are used for the fit by Eq.~(\ref{eq:velocity_profile}).
Figure \ref{fig:sim_result}(e) shows all of the profile data, where $v_{\rm t}$ and $z'$ are scaled by $v_{\rm t0}$ and $h_{\rm s}$.
All of the profile data are collapsed into a single curve, which support the validity of an exponential profile (Eq.~(\ref{eq:velocity_profile})).

Next, according to the experimental result (Fig.~\ref{fig:exp_result}(c)), $v_{\rm t0}$ is a function of $|\nabla h|$, and can be fitted by Eq.~(\ref{eq:NDT_model_surface}).
Figure~\ref{fig:sim_result}(f) shows the relation between $v_{\rm t0}$ and $|\nabla h|$ obtained by DEM simulations.
Although the data particularly at early $t$ scatter compared to Fig.~\ref{fig:sim_result}(a), the NDT model for the surface flow (Eq.~(\ref{eq:NDT_model_surface})) almost holds, suggesting the consistency with the experimental result.
The best fitted $c_{\rm 0}$ values are also plotted in Fig.~\ref{fig:sim_result}(c), which are approximately twice as large as $c$ (Eq.~(\ref{eq:c_csurf})) and do not depend on $e$.
This relation also agrees with the experimental result (Eq.~(\ref{eq:c_csurf})).

Finally, the relation between $h_{\rm s}$ and $|\nabla h|$ is shown in Fig.~\ref{fig:sim_result}(g), which suggests complex dependence.
$h_{\rm s}$ is a function of not only $|\nabla h|$ but also the analysis position $r$ that can be read by $h(r)$.
However, we empirically found that $h_{\rm s}/h$ is almost scaled by only $|\nabla h|$ as shown in Fig.~\ref{fig:sim_result}(h), where a linear dependence can be observed: 
\begin{equation}
\frac{h_{\rm s}}{h}=a|\nabla h|,
\label{eq:h0_h}
\end{equation}
where $a$ is a constant, which is estimated as $a=2.41$ by means of the linear least-squares method.
This trend means that the shear band gets enhanced as increasing the inclination.
When $|\nabla h|$ is fixed, Figs.~\ref{fig:sim_result}(g) and (h) suggest that $h_{\rm s}$ increases with $h$, but the ratio $h_{\rm s}/h$ is almost constant independent of $h$.
The inset of Fig.~\ref{fig:sim_result}(h) shows the analysis results obtained in various vibration conditions, where $h_{\rm s}/h$ is little influenced.

\subsection{Dependence on system size}
\label{sec:Dependence_on_system_size}

From these characteristics of the shear band structure (Eq.~(\ref{eq:h0_h}) and Fig.~\ref{fig:sim_result}(h)), it can be anticipated that the result is not influenced by the system size $R$ or the ratio $R/d$. 
In fact, the NDT model holds with the same value of $c$ when $R/d$ is varied by changing grain diameters ($d=0.5$ and $1.0$~mm) under the fixed system size ($R=40$~mm) in the experiment~\cite{Tsuji2018}.
In contrast to this, here DEM simulations attempt to expand the understanding of the dependence on the system size by directly changing $R$.

\begin{figure}[t]
\begin{center}
\includegraphics[width=80mm]{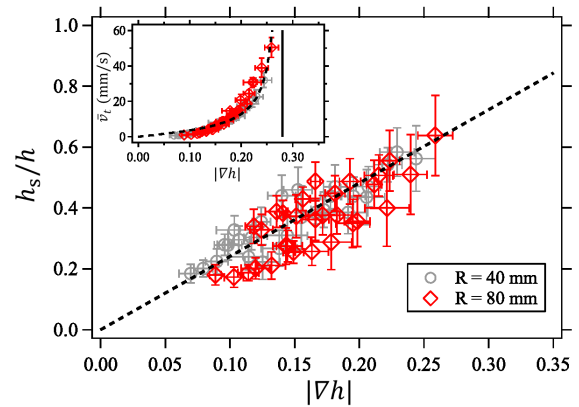}
\end{center}
\caption{
Comparison of the data obtained under different system size $R=40$ and $80$ mm.
The diameter of grains is fixed as $d=0.5$~mm.
The main panel shows the scaled thickness of the shear band $h_s/h$ as a function of slope $|\nabla h|$, while the relation between depth-averaged velocity $\bar{v}_{\rm t}$ and slope $|\nabla h|$ is plotted in the inset.
The black dashed curves and solid line represent the same ones used in Figs.~\ref{fig:sim_result}(a) and (h).
}
\label{fig:R320}
\end{figure}

The same DEM simulations as explained in Sec.~\ref{sec:Numerical simulation} are conducted under a larger plate $R=80$~mm with grains of $d=0.5$~mm.
The experimental conditions are $A=0.04$~mm and $f=200$~Hz $(\Gamma=6)$, which are the same as those used in Fig.~\ref{fig:sim_result}(h).
The velocity profile ${v}_{\rm t}(z')$ is measured at three positions ($r=20, 40~{\rm and}~60~{\rm mm}$) and various time $t$, and the characteristic thickness of the shear band $h_s$ is estimated for each data as well. 
The result is shown in Fig.~\ref{fig:R320}, which is compared to the data obtained with $R=40$~mm (same data as Fig.~\ref{fig:sim_result}(h)).
As expected, identical shear band structure expressed by Eq.~(\ref{eq:h0_h}) can be observed even if we change $R$ or $R/d$.  
In addition, as shown in the inset of Fig.~\ref{fig:R320}, it has been confirmed that the bulk flow property characterized by $\bar{v}_{\rm t}$ is independent of $R$ or $R/d$, which is also consistent with~Ref.~\cite{Tsuji2018}.

\section{Discussion}
\label{sec:Discussion}

\subsection{Consistency with the NDT model}
\label{subsec:Consistency_to_the_NDT_model}

Since DEM simulations have revealed the particle behavior inside the pile so far, the NDT model can be derived by integrating the velocity profile along the vertical direction.
Using Eqs.~(\ref{eq:depth_averaged0}), (\ref{eq:velocity_profile}) and (\ref{eq:h0_h}), the depth-averaged velocity $\bar{v}_{\rm t}$ can be computed as
\begin{eqnarray}
\bar{v}_{\rm t}&=&\frac{1}{h}\int_0^h v_{\rm t}(z')dz' \nonumber\\
&=& \frac{1}{h}\int_0^h v_{\rm t0} \exp \left( -\frac{z'}{h_{\rm s}}\right)dz' \nonumber\\
&=& f(|\nabla h|) \frac{c_0v_{\rm vib}}{\mu^2}\frac{|\nabla h|}{1-(|\nabla h|/\mu)^2}, 
\label{eq:depth_averaged1}
\end{eqnarray}
where
\begin{eqnarray}
f(|\nabla h|) = a|\nabla h| \left\{ 1-\exp \left(-\frac{1}{a|\nabla h|} \right)  \right\}.
\label{eq:depth_averaged2}
\end{eqnarray}
Although the dependence on $h$ does not appear in $\bar{v}_{\rm t}$ as predicted by the NDT model, Eq.~(\ref{eq:depth_averaged1}) is not equal to Eq.~(\ref{eq:NDT_model}).
However, as reported by Ref.~\cite{Tsuji2018}, the NDT model does not hold in smaller $|\nabla h|$ ranges, where the relaxation almost halts leaving finite slopes.
In fact, Figs.~\ref{fig:exp_result}(a) and \ref{fig:sim_result}(a) show that the model curves exhibit misfits to the data as the slopes approach zero.
With respect to this reason, it can be speculated that the inertial energy supplied to grains is insufficient to overcome potential barriers of their neighbors in small $|\nabla h|$ ranges~\cite{Jaeger1989, Roering2004}.
On the other hand, this fact suggests that the NDT model is mainly suitable to predict the bulk flow property in a large $|\nabla h|$ range, where non-linearity clearly appears and the relaxation dynamics is practically dominated.
We therefore focus on how Eq.~(\ref{eq:depth_averaged2}) behaves in the vicinity of a divergence point ($|\nabla h| \simeq \mu$).

To this end, we introduce a variable $\epsilon = \mu - |\nabla h|$, and Eq.~(\ref{eq:depth_averaged2}) is subjected to variable conversion:
\begin{eqnarray}
f(\epsilon) = a(\mu - \epsilon) \left\{ 1-\exp \left(-\frac{1}{a(\mu-\epsilon)} \right)  \right\}.
\label{eq:depth_averaged3}
\end{eqnarray}
In order to observe the behavior of this function around $\epsilon \to 0~(|\nabla h| \to \mu)$, Taylor expansion is conducted as follows:
\begin{eqnarray}
f(\epsilon) &=& a\mu \left\{ 1-\exp \left(-\frac{1}{a\mu} \right) \right\} + \nonumber\\
&& \epsilon\left\{ \left( a + \frac{1}{\mu}\right) \exp \left( -\frac{1}{a\mu}\right)  - a \right\} + o(\epsilon^2) \nonumber\\
&\sim& 0.5 - \epsilon,
\label{eq:depth_averaged4}
\end{eqnarray}
where $a=2.41$ and $\mu=0.28$ are substituted.
Therefore, in the limit of $\epsilon \to 0$, which is important for the fitting of the NDT model, Eq.~(\ref{eq:depth_averaged1}) can be read as
\begin{eqnarray}
\bar{v}_{\rm t} \sim \frac{0.5c_0v_{\rm vib}}{\mu^2}\frac{|\nabla h|}{1-(|\nabla h|/\mu)^2}.
\label{eq:depth_averaged5}
\end{eqnarray}
Since Eq.~(\ref{eq:NDT_model}) is identical to Eq.~(\ref{eq:depth_averaged5}) with Eq.~(\ref{eq:c_csurf}), the NDT model has been consistently reproduced from the integration of the velocity profile obtained by DEM simulations.
For small $|\nabla h|$ regimes, $c$ should decrease linearly as shown in Fig.~\ref{fig:sim_result}(h) and Eq.~(\ref{eq:depth_averaged4}).
However, its effect is rather limited to discuss the practical relaxation of the pile.

\begin{figure}[b]
\begin{center}
\includegraphics[width=60mm]{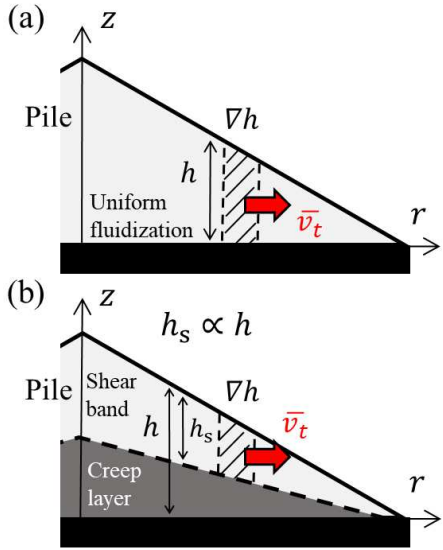}
\end{center}
\caption{
Schematic images of (a) the derivation of the NDT model proposed by Ref.~\cite{Tsuji2018} and (b) the $h$-proportional shear band structure revealed by DEM simulations.
}
\label{fig:model_image}
\end{figure}

Another remarkable point revealed by DEM simulations is that heap flows create shear band structure.
In other words, clearly fluidized regimes are localized around the surface with thickness $h_{\rm s}$, below which creeping flows exhibit.
This fact is in contrast to the assumption used in the derivation of the NDT model~\cite{Tsuji2018} that the whole pile is uniformly fluidized as illustrated in Fig.~\ref{fig:model_image}(a).
Actually, however, $h_{\rm s}$ is proportional to $h$ (Eq.~(\ref{eq:h0_h})) when $|\nabla h|$ is fixed, which leads to a true image of the flow drawn in Fig.~\ref{fig:model_image}(b).
Beside, Figs.~\ref{fig:sim_result}(e) and \ref{fig:R320} imply that the velocity profile is similar at any position and time independent of heap size as long as the vibration is strong enough to mobilize the whole granular pile.
This characteristic differs from a conventional shear band structure with a constant thickness everywhere as observed in heap flows in the absence of vibration~(e.g., \cite{Katsuragi2010}).
Although the detailed pictures of Figs.~\ref{fig:model_image}(a) and (b) are different from each other, $h$ determines the fluidized thickness $h_{\rm s}$ in both cases. 
This is why $\bar{v}_{\rm t} = q/h (\propto q/h_{\rm s})$ can be described as a function of only $|\nabla h|$, and the NDT model derived on a basis of Fig.~\ref{fig:model_image}(a) is still valid for the granular flow consisting of peculiar shear band structure as shown in Fig.~\ref{fig:model_image}(b).

\subsection{Potential applicability}
\label{subsec:Potential_applicability}

It is also noteworthy that sediment transport from soil-mantled hillslopes shows a similar nonlinear property. 
The relation between sediment flux and hillslope gradient exhibits nonlinearity like Eq.~(\ref{eq:NDT_model}).
In other words, the flux increases divergently as the slope approaches a certain critical slope, which is reported by field observations~\cite{Roering1999} and field measurements~\cite{Gabet2003}, where environmental disturbance (e.g., earthquakes, rainsplash and biogenic activity) is considered to mobilize regolith particles.
This natural process is mimicked by granular flows with acoustic noise in laboratory experiments~\cite{Roering2001, Roering2004, Furbish2008}; and with random perturbation in DEM simulations~\cite{BenDror2018,Ferdowsi2018}.
Although the applied perturbation types of these studies are different from mechanically-controlled vibration used in our study, similar nonlinear transport properties are reported. 
From this similarity, it can be expected that the framework of our modeling for heap flows in the presence of vibration will be potentially applicable to other experimental configurations with different disturbance types.

\subsection{Limitations and future works}
\label{subsec:Limitations_and_future_works}

Finally, several limitations of the model are discussed here.
Although these limitations introduced below are beyond the scope of the present paper, they are important open issues left for future works.

The first limitation is that the vibration range where the NDT model can be applied is limited.
This study focuses on the vibration conditions of $v_{\rm vib}=10-200$~mm/s and $\Gamma=2-10$.
When increasing the vibration strength above this range, the transition into a granular-gas phase~\cite{Jaeger1996} will occur, where the NDT model is no longer suitable.
Conversely, as approaching a critical fluidization condition $\Gamma=\Gamma_c$, the NDT model will break down at some point, where the whole layer is not fluidized, i.e., the characteristic shear band drawn in Fig.~\ref{fig:model_image}(b) will not be created.
The critical conditions to distinguish these multiple regimes will need to be investigated.

The second limitation is that the modeling in this study completely neglects the contribution of velocity fluctuation.
Since the energy is transfered through a vibrating disk, the boundary conditions, such as those proposed in Ref.~\cite{Richman1993}, should be satisfied, which also enables us to calculate a profile of granular temperature in the system.
According to a kinetic theory for dense fluidized flows (e.g., \cite{Jenkins2012}), it is predicted that the viscosity, which connects the shear stress with shear rate, changes as a function of local granular temperature.
Therefore, taking velocity fluctuation into consideration would be important to theoretically explain the velocity profile which should be governed by viscosity.

The third limitation is that the bulk frictional coefficient $\mu$, which is defined as the ratio of the shear stress to the pressure, is assumed to be constant.
The dependence of $\mu$ on vibration conditions has been investigated by experiments~\cite{Sanchez2007} and DEM simulations~\cite{Khefif2018}.
These studies observe a spreading granular droplet under horizontal vibration, and report that $\mu$ varies as a function of an inertial dimensionless parameter $I=v_{\rm vib}/\sqrt{gd}$, or the square root of a shaking parameter $S={v_{\rm vib}}^2/gd$~\cite{Pak1993}.
From this, one might speculate that the dynamics of heap flows under vibration can be discussed on the analogy of local rheology~\cite{Jop2005,Jop2006}.

The last important open question is ``what underlying nature determines $c$?''
Since the value of $c$ is much less than $1$, most of the inputted energy is not used for the bulk granular transport.
Moreover, strictly speaking, the value of $c$ (Fig.~\ref{fig:exp_result}) shows a slight upward trend with $v_{\rm vib}$ when $f$ is fixed, although $c$ can be regarded as a constant approximately.
The reason for this could be related to some missing factors described above.
In any case, to solve this issue it is necessary to fully evaluate the energy partition among dissipation (inelasticity and friction), random motion (granular temperature), and collective motion (mean flow determining the value of $c$).

\section{Conclusion}
\label{sec:Conclusion}

For the purpose of understanding the granular heap flow on a pile fully-fluidized by relatively strong vibration, we have experimentally and numerically studied the relaxation dynamics of a granular pile on a vertically-vibrating plate.
To explain the relation between the depth-averaged velocity and local slope, the NDT model (Eq.~(\ref{eq:NDT_model})) has been proposed in Ref.~\cite{Tsuji2018}, which turns out to be applicable to the surface velocity (Eq.~(\ref{eq:NDT_model_surface})) as well.
These results are also satisfied in the DEM simulations, which support the universality of the modeling.
The comparison of the model fitting parameters $c$ and $c_0$, which are constant independent of both experimental and numerical conditions, suggests that the surface velocity is approximately twice as large as the depth-averaged velocity (Eq.~(\ref{eq:c_csurf})).
This result predicts that the flow velocity decreases as going deeper from the surface, which has been confirmed by measuring the internal velocity profiles obtained by DEM simulations.
Moreover, it has been revealed that the relaxing pile creates shear band around the surface with exponentially-decreasing velocity profile (Eq.~(\ref{eq:velocity_profile})).
Its characteristic thickness, however, is not constant but proportional to the local height of the pile.
We have also confirmed that these flow properties are independent of the system size.
Finally, by integrating the exponential velocity function with this peculiar shear band structure from the base to the surface, the depth-averaged velocity described by the NDT model can be successfully deduced.
Although these results are mostly based on empirical findings for now, the bulk transport picture proposed by Ref.~\cite{Tsuji2018} has been consistently bridged to the particle-scale detailed picture revealed by this study.

\section*{Acknowledgments}
This study has been supported by JSPS KAKENHI Grants No. 15H03707, No. 16H04025, No. 17H05420, No. 17J05552, and No. 18H03679.

\appendix

\section{Algorithm of the profile pattern matching}
\label{appendixA}

\begin{figure}[b]
\begin{center}
\includegraphics[width=60mm]{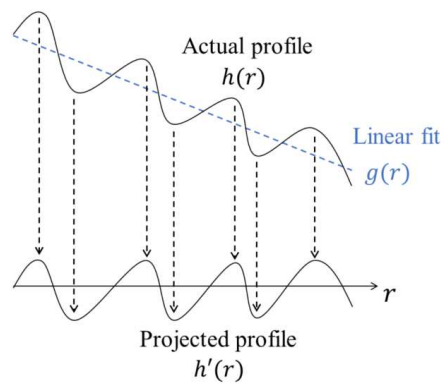}
\end{center}
\caption{Schematic image of the projection of a profile onto the horizontal axis.}
\label{fig:projected_profile}
\end{figure}

Here the way of estimating $\Delta r$ is explained.
Since $\Delta r$ is defined as the $r$ component of the distance that a profile moves during $\Delta t$, firstly a profile, which is actually inclined along slope, is projected onto the horizontal axis.
To this end, as illustrated in Fig.~\ref{fig:projected_profile}, a profile $h(r)$ is fitted by a linear function ($g(r)=a_0+a_1 r$, where $a_0$ and $a_1$ are fitting parameters), and then the profile along the $r$ axis $h'(r)$ is created as
\begin{equation}
h'(r) = h(r)-g(r).
\label{eq:mapping}
\end{equation}
$\Delta r$ can be determined by finding the position where one dimensional cross correlation function $f_{cc}(\delta r)$ between two profiles $h'(r,t)$ and $h'(r,t+\Delta t)$ shows the maximum value. 
The cross correlation function is defined as
\begin{equation}
f_{cc}(\delta r) = \frac{1}{H'(t)}\int_{r_s}^{r_e} h'(r,t)  h'(r+\delta r, t+\Delta t) dr,
\label{eq:cross_correlation_fn}
\end{equation}
where $r_s$ and $r_e$ represent the start- and end-points of the spacial window considered for the analysis, and the normalization term $H'(t)$ is given by
\begin{equation}
H'(t)=\int_{r_s}^{r_e} h'(r,t)^2 dr.
\label{eq:cross_correlation_fn2}
\end{equation}
Here, $r_e - r_s=10~{\rm mm}$ is employed as explained in Sec.~\ref{sec:Surface velocity}.
Since the correlation function is normalized by ${H'(t)}$ in Eq.~\eqref{eq:cross_correlation_fn}, $f_{cc}(\delta r)=1$ corresponds to the complete match of two profiles.
By changing $\delta r$ systematically, $\Delta r$ can be estimated, where $f_{cc}(\Delta r)$ shows the peak value of the correlation function.
In this study, six different $\Delta t$ are applied to obtain reliable data, which increase consecutively at intervals of the temporal resolution of the profile measurement by the laser ($0.02$~s).
Hereafter, six different $\Delta t$ are denoted by $\Delta t_i$, and their corresponding $\Delta r$ are expressed as $\Delta r_i$, where $i=1,2,\cdots,6$.
$\Delta t_i$ is given by $\Delta t_0 + 0.02i~{\rm s}$, where a constant $\Delta t_0$ depends on the analysis position $r$.
In general, the surface velocity increases with $r$ as a local slope gets steeper.
Therefore, it is preferable for the accurate measurement that smaller $\Delta t_0$ is chosen for larger $r$ ranges.
Table~\ref{tab:list} shows the list of $\Delta t_0$ for various analysis positions $(r=5,15,25,35~{\rm mm})$.
These values are chosen in order for the typical value of $\Delta r$ to be a few grain diameter.

\begin{table}[b]
\setlength{\tabcolsep}{3mm}
\begin{center}
\begin{tabular}{l|cccc}
\hline
$r$ (mm) &  $5$     &	$15$	& $25$ 	& $35$	 \\
\hline
$\Delta t_0$ (s)    &  $0.12$   &	$0.08$	& $0.04$  & $0.00$	 \\	
\hline
\end{tabular}
\caption{List of $\Delta t_0$ for various analysis positions $r$.}
\label{tab:list}
\end{center}
\end{table}

Finally, the surface velocity $v_{\rm t0}$ is estimated using the weighted average:
\begin{equation}
v_{\rm t0} = \sum_{i=1}^6 \left( \frac{\Delta r_i}{\Delta t_i} \cdot \frac{f_{cc}(\Delta r_i)^2}{f_{cc}^{tot}} \right),
\label{eq:surface_vel}
\end{equation}
where $f_{cc}^{tot} = \sum_{i=1}^6 f_{cc}(\Delta r_i)^2 $.
Note that only the velocity with $f_{cc}(\Delta r_i)>0.4$ is used to calculate $v_{\rm t0}$ so that the data obtained by low-correlated matching, which could be inaccurate, can be removed. 
We have confirmed that the result does not change once this threshold value becomes larger than $0.4$.

\newpage

\bibliography{sandpile}

\begin{thebibliography}{51}%
\makeatletter
\providecommand \@ifxundefined [1]{%
 \@ifx{#1\undefined}
}%
\providecommand \@ifnum [1]{%
 \ifnum #1\expandafter \@firstoftwo
 \else \expandafter \@secondoftwo
 \fi
}%
\providecommand \@ifx [1]{%
 \ifx #1\expandafter \@firstoftwo
 \else \expandafter \@secondoftwo
 \fi
}%
\providecommand \natexlab [1]{#1}%
\providecommand \enquote  [1]{``#1''}%
\providecommand \bibnamefont  [1]{#1}%
\providecommand \bibfnamefont [1]{#1}%
\providecommand \citenamefont [1]{#1}%
\providecommand \href@noop [0]{\@secondoftwo}%
\providecommand \href [0]{\begingroup \@sanitize@url \@href}%
\providecommand \@href[1]{\@@startlink{#1}\@@href}%
\providecommand \@@href[1]{\endgroup#1\@@endlink}%
\providecommand \@sanitize@url [0]{\catcode `\\12\catcode `\$12\catcode
  `\&12\catcode `\#12\catcode `\^12\catcode `\_12\catcode `\%12\relax}%
\providecommand \@@startlink[1]{}%
\providecommand \@@endlink[0]{}%
\providecommand \url  [0]{\begingroup\@sanitize@url \@url }%
\providecommand \@url [1]{\endgroup\@href {#1}{\urlprefix }}%
\providecommand \urlprefix  [0]{URL }%
\providecommand \Eprint [0]{\href }%
\providecommand \doibase [0]{http://dx.doi.org/}%
\providecommand \selectlanguage [0]{\@gobble}%
\providecommand \bibinfo  [0]{\@secondoftwo}%
\providecommand \bibfield  [0]{\@secondoftwo}%
\providecommand \translation [1]{[#1]}%
\providecommand \BibitemOpen [0]{}%
\providecommand \bibitemStop [0]{}%
\providecommand \bibitemNoStop [0]{.\EOS\space}%
\providecommand \EOS [0]{\spacefactor3000\relax}%
\providecommand \BibitemShut  [1]{\csname bibitem#1\endcsname}%
\let\auto@bib@innerbib\@empty
%</preamble>
\bibitem [{\citenamefont {Lemieux}\ and\ \citenamefont
  {Durian}(2000)}]{Lemieux2000}%
  \BibitemOpen
  \bibfield  {author} {\bibinfo {author} {\bibfnamefont {P.-A.}\ \bibnamefont
  {Lemieux}}\ and\ \bibinfo {author} {\bibfnamefont {D.~J.}\ \bibnamefont
  {Durian}},\ }\href@noop {} {\bibfield  {journal} {\bibinfo  {journal} {Phys.
  Rev. Lett.}\ }\textbf {\bibinfo {volume} {85}},\ \bibinfo {pages} {4273}
  (\bibinfo {year} {2000})}\BibitemShut {NoStop}%
\bibitem [{\citenamefont {Komatsu}\ \emph {et~al.}(2001)\citenamefont
  {Komatsu}, \citenamefont {Inagaki}, \citenamefont {Nakagawa},\ and\
  \citenamefont {Nasuno}}]{Komatsu2001}%
  \BibitemOpen
  \bibfield  {author} {\bibinfo {author} {\bibfnamefont {T.~S.}\ \bibnamefont
  {Komatsu}}, \bibinfo {author} {\bibfnamefont {S.}~\bibnamefont {Inagaki}},
  \bibinfo {author} {\bibfnamefont {N.}~\bibnamefont {Nakagawa}}, \ and\
  \bibinfo {author} {\bibfnamefont {S.}~\bibnamefont {Nasuno}},\ }\href@noop {}
  {\bibfield  {journal} {\bibinfo  {journal} {Phys. Rev. Lett.}\ }\textbf
  {\bibinfo {volume} {86}},\ \bibinfo {pages} {1757} (\bibinfo {year}
  {2001})}\BibitemShut {NoStop}%
\bibitem [{\citenamefont {{GDR MiDi}}(2004)}]{GDR2004}%
  \BibitemOpen
  \bibfield  {author} {\bibinfo {author} {\bibnamefont {{GDR MiDi}}},\
  }\href@noop {} {\bibfield  {journal} {\bibinfo  {journal} {Eur. Phys. J. E}\
  }\textbf {\bibinfo {volume} {14}},\ \bibinfo {pages} {341} (\bibinfo {year}
  {2004})}\BibitemShut {NoStop}%
\bibitem [{\citenamefont {Jop}\ \emph {et~al.}(2005)\citenamefont {Jop},
  \citenamefont {Forterre},\ and\ \citenamefont {Pouliquen}}]{Jop2005}%
  \BibitemOpen
  \bibfield  {author} {\bibinfo {author} {\bibfnamefont {P.}~\bibnamefont
  {Jop}}, \bibinfo {author} {\bibfnamefont {Y.}~\bibnamefont {Forterre}}, \
  and\ \bibinfo {author} {\bibfnamefont {O.}~\bibnamefont {Pouliquen}},\
  }\href@noop {} {\bibfield  {journal} {\bibinfo  {journal} {J. Fluid Mech.}\
  }\textbf {\bibinfo {volume} {541}},\ \bibinfo {pages} {167} (\bibinfo {year}
  {2005})}\BibitemShut {NoStop}%
\bibitem [{\citenamefont {Jop}\ \emph {et~al.}(2006)\citenamefont {Jop},
  \citenamefont {Forterre},\ and\ \citenamefont {Pouliquen}}]{Jop2006}%
  \BibitemOpen
  \bibfield  {author} {\bibinfo {author} {\bibfnamefont {P.}~\bibnamefont
  {Jop}}, \bibinfo {author} {\bibfnamefont {Y.}~\bibnamefont {Forterre}}, \
  and\ \bibinfo {author} {\bibfnamefont {O.}~\bibnamefont {Pouliquen}},\
  }\href@noop {} {\bibfield  {journal} {\bibinfo  {journal} {Nature}\ }\textbf
  {\bibinfo {volume} {441}},\ \bibinfo {pages} {727} (\bibinfo {year}
  {2006})}\BibitemShut {NoStop}%
\bibitem [{\citenamefont {Katsuragi}\ \emph {et~al.}(2010)\citenamefont
  {Katsuragi}, \citenamefont {Abate},\ and\ \citenamefont
  {Durian}}]{Katsuragi2010}%
  \BibitemOpen
  \bibfield  {author} {\bibinfo {author} {\bibfnamefont {H.}~\bibnamefont
  {Katsuragi}}, \bibinfo {author} {\bibfnamefont {A.~R.}\ \bibnamefont
  {Abate}}, \ and\ \bibinfo {author} {\bibfnamefont {D.~J.}\ \bibnamefont
  {Durian}},\ }\href@noop {} {\bibfield  {journal} {\bibinfo  {journal} {Soft
  Matter}\ }\textbf {\bibinfo {volume} {6}},\ \bibinfo {pages} {3023} (\bibinfo
  {year} {2010})}\BibitemShut {NoStop}%
\bibitem [{\citenamefont {GRAY}(2001)}]{Gray2001}%
  \BibitemOpen
  \bibfield  {author} {\bibinfo {author} {\bibfnamefont {J.~M. N.~T.}\
  \bibnamefont {GRAY}},\ }\href@noop {} {\bibfield  {journal} {\bibinfo
  {journal} {J. Fluid Mech.}\ }\textbf {\bibinfo {volume} {441}},\ \bibinfo
  {pages} {1} (\bibinfo {year} {2001})}\BibitemShut {NoStop}%
\bibitem [{\citenamefont {Bonamy}(2002)}]{Bonamy2002}%
  \BibitemOpen
  \bibfield  {author} {\bibinfo {author} {\bibfnamefont {D.}~\bibnamefont
  {Bonamy}},\ }\href@noop {} {\bibfield  {journal} {\bibinfo  {journal} {Phys.
  Fluids}\ }\textbf {\bibinfo {volume} {14}},\ \bibinfo {pages} {1666}
  (\bibinfo {year} {2002})}\BibitemShut {NoStop}%
\bibitem [{\citenamefont {Courrech~du Pont}\ \emph {et~al.}(2005)\citenamefont
  {Courrech~du Pont}, \citenamefont {Fischer}, \citenamefont {Gondret},
  \citenamefont {Perrin},\ and\ \citenamefont {Rabaud}}]{Pont2005}%
  \BibitemOpen
  \bibfield  {author} {\bibinfo {author} {\bibfnamefont {S.}~\bibnamefont
  {Courrech~du Pont}}, \bibinfo {author} {\bibfnamefont {R.}~\bibnamefont
  {Fischer}}, \bibinfo {author} {\bibfnamefont {P.}~\bibnamefont {Gondret}},
  \bibinfo {author} {\bibfnamefont {B.}~\bibnamefont {Perrin}}, \ and\ \bibinfo
  {author} {\bibfnamefont {M.}~\bibnamefont {Rabaud}},\ }\href@noop {}
  {\bibfield  {journal} {\bibinfo  {journal} {Phys. Rev. Lett.}\ }\textbf
  {\bibinfo {volume} {94}},\ \bibinfo {pages} {048003} (\bibinfo {year}
  {2005})}\BibitemShut {NoStop}%
\bibitem [{\citenamefont {Yang}\ \emph {et~al.}(2008)\citenamefont {Yang},
  \citenamefont {Yu}, \citenamefont {McElroy},\ and\ \citenamefont
  {Bao}}]{Yang2008}%
  \BibitemOpen
  \bibfield  {author} {\bibinfo {author} {\bibfnamefont {R.}~\bibnamefont
  {Yang}}, \bibinfo {author} {\bibfnamefont {A.}~\bibnamefont {Yu}}, \bibinfo
  {author} {\bibfnamefont {L.}~\bibnamefont {McElroy}}, \ and\ \bibinfo
  {author} {\bibfnamefont {J.}~\bibnamefont {Bao}},\ }\href@noop {} {\bibfield
  {journal} {\bibinfo  {journal} {Powder Technol.}\ }\textbf {\bibinfo {volume}
  {188}},\ \bibinfo {pages} {170 } (\bibinfo {year} {2008})}\BibitemShut
  {NoStop}%
\bibitem [{\citenamefont {Kleinhans}\ \emph {et~al.}(2011)\citenamefont
  {Kleinhans}, \citenamefont {Markies}, \citenamefont {de~Vet}, \citenamefont
  {in~'t Veld},\ and\ \citenamefont {Postema}}]{Kleinhans2011}%
  \BibitemOpen
  \bibfield  {author} {\bibinfo {author} {\bibfnamefont {M.~G.}\ \bibnamefont
  {Kleinhans}}, \bibinfo {author} {\bibfnamefont {H.}~\bibnamefont {Markies}},
  \bibinfo {author} {\bibfnamefont {S.~J.}\ \bibnamefont {de~Vet}}, \bibinfo
  {author} {\bibfnamefont {A.~C.}\ \bibnamefont {in~'t Veld}}, \ and\ \bibinfo
  {author} {\bibfnamefont {F.~N.}\ \bibnamefont {Postema}},\ }\href@noop {}
  {\bibfield  {journal} {\bibinfo  {journal} {Journal of Geophysical Research:
  Planets}\ }\textbf {\bibinfo {volume} {116}},\ \bibinfo {pages} {E11004}
  (\bibinfo {year} {2011})}\BibitemShut {NoStop}%
\bibitem [{\citenamefont {Amon}\ \emph {et~al.}(2013)\citenamefont {Amon},
  \citenamefont {Niculescu},\ and\ \citenamefont {Utter}}]{Amon2013}%
  \BibitemOpen
  \bibfield  {author} {\bibinfo {author} {\bibfnamefont {D.~L.}\ \bibnamefont
  {Amon}}, \bibinfo {author} {\bibfnamefont {T.}~\bibnamefont {Niculescu}}, \
  and\ \bibinfo {author} {\bibfnamefont {B.~C.}\ \bibnamefont {Utter}},\
  }\href@noop {} {\bibfield  {journal} {\bibinfo  {journal} {Phys. Rev. E}\
  }\textbf {\bibinfo {volume} {88}},\ \bibinfo {pages} {012203} (\bibinfo
  {year} {2013})}\BibitemShut {NoStop}%
\bibitem [{\citenamefont {Swisher}\ and\ \citenamefont
  {Utter}(2014)}]{Swisher2014}%
  \BibitemOpen
  \bibfield  {author} {\bibinfo {author} {\bibfnamefont {N.~C.}\ \bibnamefont
  {Swisher}}\ and\ \bibinfo {author} {\bibfnamefont {B.~C.}\ \bibnamefont
  {Utter}},\ }\href@noop {} {\bibfield  {journal} {\bibinfo  {journal}
  {Granular Matter}\ }\textbf {\bibinfo {volume} {16}},\ \bibinfo {pages} {175}
  (\bibinfo {year} {2014})}\BibitemShut {NoStop}%
\bibitem [{\citenamefont {Yamada}\ and\ \citenamefont
  {Katsuragi}(2014)}]{Yamada2014}%
  \BibitemOpen
  \bibfield  {author} {\bibinfo {author} {\bibfnamefont {T.~M.}\ \bibnamefont
  {Yamada}}\ and\ \bibinfo {author} {\bibfnamefont {H.}~\bibnamefont
  {Katsuragi}},\ }\href@noop {} {\bibfield  {journal} {\bibinfo  {journal}
  {Planet. and Space Sci.}\ }\textbf {\bibinfo {volume} {100}},\ \bibinfo
  {pages} {79} (\bibinfo {year} {2014})}\BibitemShut {NoStop}%
\bibitem [{\citenamefont {Breu}\ \emph {et~al.}(2003)\citenamefont {Breu},
  \citenamefont {Ensner}, \citenamefont {Kruelle},\ and\ \citenamefont
  {Rehberg}}]{Breu2003}%
  \BibitemOpen
  \bibfield  {author} {\bibinfo {author} {\bibfnamefont {A.~P.~J.}\
  \bibnamefont {Breu}}, \bibinfo {author} {\bibfnamefont {H.-M.}\ \bibnamefont
  {Ensner}}, \bibinfo {author} {\bibfnamefont {C.~A.}\ \bibnamefont {Kruelle}},
  \ and\ \bibinfo {author} {\bibfnamefont {I.}~\bibnamefont {Rehberg}},\
  }\href@noop {} {\bibfield  {journal} {\bibinfo  {journal} {Phys. Rev. Lett.}\
  }\textbf {\bibinfo {volume} {90}},\ \bibinfo {pages} {014302} (\bibinfo
  {year} {2003})}\BibitemShut {NoStop}%
\bibitem [{\citenamefont {Iikawa}\ \emph {et~al.}(2015)\citenamefont {Iikawa},
  \citenamefont {Bandi},\ and\ \citenamefont {Katsuragi}}]{Iikawa2015}%
  \BibitemOpen
  \bibfield  {author} {\bibinfo {author} {\bibfnamefont {N.}~\bibnamefont
  {Iikawa}}, \bibinfo {author} {\bibfnamefont {M.~M.}\ \bibnamefont {Bandi}}, \
  and\ \bibinfo {author} {\bibfnamefont {H.}~\bibnamefont {Katsuragi}},\
  }\href@noop {} {\bibfield  {journal} {\bibinfo  {journal} {J. Phys. Soc.
  Jpn.}\ }\textbf {\bibinfo {volume} {84}},\ \bibinfo {pages} {094401}
  (\bibinfo {year} {2015})}\BibitemShut {NoStop}%
\bibitem [{\citenamefont {Caballero-Robledo}\ and\ \citenamefont
  {Cl{\'e}ment}(2009)}]{Caballero-Robledo2009}%
  \BibitemOpen
  \bibfield  {author} {\bibinfo {author} {\bibfnamefont {G.~A.}\ \bibnamefont
  {Caballero-Robledo}}\ and\ \bibinfo {author} {\bibfnamefont {E.}~\bibnamefont
  {Cl{\'e}ment}},\ }\href@noop {} {\bibfield  {journal} {\bibinfo  {journal}
  {Eur. Phys. J. E}\ }\textbf {\bibinfo {volume} {30}},\ \bibinfo {pages} {395}
  (\bibinfo {year} {2009})}\BibitemShut {NoStop}%
\bibitem [{\citenamefont {Jaeger}\ \emph {et~al.}(1989)\citenamefont {Jaeger},
  \citenamefont {Liu},\ and\ \citenamefont {Nagel}}]{Jaeger1989}%
  \BibitemOpen
  \bibfield  {author} {\bibinfo {author} {\bibfnamefont {H.~M.}\ \bibnamefont
  {Jaeger}}, \bibinfo {author} {\bibfnamefont {C.-h.}\ \bibnamefont {Liu}}, \
  and\ \bibinfo {author} {\bibfnamefont {S.~R.}\ \bibnamefont {Nagel}},\
  }\href@noop {} {\bibfield  {journal} {\bibinfo  {journal} {Phys. Rev. Lett.}\
  }\textbf {\bibinfo {volume} {62}},\ \bibinfo {pages} {40} (\bibinfo {year}
  {1989})}\BibitemShut {NoStop}%
\bibitem [{\citenamefont {S\'anchez}\ \emph {et~al.}(2007)\citenamefont
  {S\'anchez}, \citenamefont {Raynaud}, \citenamefont {Lanuza}, \citenamefont
  {Andreotti}, \citenamefont {Cl\'ement},\ and\ \citenamefont
  {Aranson}}]{Sanchez2007}%
  \BibitemOpen
  \bibfield  {author} {\bibinfo {author} {\bibfnamefont {I.}~\bibnamefont
  {S\'anchez}}, \bibinfo {author} {\bibfnamefont {F.}~\bibnamefont {Raynaud}},
  \bibinfo {author} {\bibfnamefont {J.}~\bibnamefont {Lanuza}}, \bibinfo
  {author} {\bibfnamefont {B.}~\bibnamefont {Andreotti}}, \bibinfo {author}
  {\bibfnamefont {E.}~\bibnamefont {Cl\'ement}}, \ and\ \bibinfo {author}
  {\bibfnamefont {I.~S.}\ \bibnamefont {Aranson}},\ }\href@noop {} {\bibfield
  {journal} {\bibinfo  {journal} {Phys. Rev. E}\ }\textbf {\bibinfo {volume}
  {76}},\ \bibinfo {pages} {060301} (\bibinfo {year} {2007})}\BibitemShut
  {NoStop}%
\bibitem [{\citenamefont {Khefif}\ \emph {et~al.}(2018)\citenamefont {Khefif},
  \citenamefont {Valance},\ and\ \citenamefont {Ould-Kaddour}}]{Khefif2018}%
  \BibitemOpen
  \bibfield  {author} {\bibinfo {author} {\bibfnamefont {S.~M.}\ \bibnamefont
  {Khefif}}, \bibinfo {author} {\bibfnamefont {A.}~\bibnamefont {Valance}}, \
  and\ \bibinfo {author} {\bibfnamefont {F.}~\bibnamefont {Ould-Kaddour}},\
  }\href@noop {} {\bibfield  {journal} {\bibinfo  {journal} {Phys. Rev. E}\
  }\textbf {\bibinfo {volume} {97}},\ \bibinfo {pages} {062903} (\bibinfo
  {year} {2018})}\BibitemShut {NoStop}%
\bibitem [{\citenamefont {Gaudel}\ and\ \citenamefont {Kiesgen~de
  Richter}(2018)}]{Gaudel2018}%
  \BibitemOpen
  \bibfield  {author} {\bibinfo {author} {\bibfnamefont {N.}~\bibnamefont
  {Gaudel}}\ and\ \bibinfo {author} {\bibfnamefont {S.}~\bibnamefont
  {Kiesgen~de Richter}},\ }\href@noop {} {\bibfield  {journal} {\bibinfo
  {journal} {Soft Matter}\ }\textbf {\bibinfo {volume} {14}},\ \bibinfo {pages}
  {9445} (\bibinfo {year} {2018})}\BibitemShut {NoStop}%
\bibitem [{\citenamefont {{Richardson Jr.}}\ \emph {et~al.}(2005)\citenamefont
  {{Richardson Jr.}}, \citenamefont {Melosh}, \citenamefont {Greenberg},\ and\
  \citenamefont {O'Brien}}]{Richardson2005}%
  \BibitemOpen
  \bibfield  {author} {\bibinfo {author} {\bibfnamefont {J.~E.}\ \bibnamefont
  {{Richardson Jr.}}}, \bibinfo {author} {\bibfnamefont {H.~J.}\ \bibnamefont
  {Melosh}}, \bibinfo {author} {\bibfnamefont {R.~J.}\ \bibnamefont
  {Greenberg}}, \ and\ \bibinfo {author} {\bibfnamefont {D.~P.}\ \bibnamefont
  {O'Brien}},\ }\href@noop {} {\bibfield  {journal} {\bibinfo  {journal}
  {Icarus}\ }\textbf {\bibinfo {volume} {179}},\ \bibinfo {pages} {325 }
  (\bibinfo {year} {2005})}\BibitemShut {NoStop}%
\bibitem [{\citenamefont {Michel}\ \emph {et~al.}(2009)\citenamefont {Michel},
  \citenamefont {O'Brien}, \citenamefont {Abe},\ and\ \citenamefont
  {Hirata}}]{Michel2009}%
  \BibitemOpen
  \bibfield  {author} {\bibinfo {author} {\bibfnamefont {P.}~\bibnamefont
  {Michel}}, \bibinfo {author} {\bibfnamefont {D.}~\bibnamefont {O'Brien}},
  \bibinfo {author} {\bibfnamefont {S.}~\bibnamefont {Abe}}, \ and\ \bibinfo
  {author} {\bibfnamefont {N.}~\bibnamefont {Hirata}},\ }\href@noop {}
  {\bibfield  {journal} {\bibinfo  {journal} {Icarus}\ }\textbf {\bibinfo
  {volume} {200}},\ \bibinfo {pages} {503 } (\bibinfo {year}
  {2009})}\BibitemShut {NoStop}%
\bibitem [{\citenamefont {Evesque}\ and\ \citenamefont
  {Rajchenbach}(1989)}]{Evesque1989}%
  \BibitemOpen
  \bibfield  {author} {\bibinfo {author} {\bibfnamefont {P.}~\bibnamefont
  {Evesque}}\ and\ \bibinfo {author} {\bibfnamefont {J.}~\bibnamefont
  {Rajchenbach}},\ }\href@noop {} {\bibfield  {journal} {\bibinfo  {journal}
  {Phys. Rev. Lett.}\ }\textbf {\bibinfo {volume} {62}},\ \bibinfo {pages} {44}
  (\bibinfo {year} {1989})}\BibitemShut {NoStop}%
\bibitem [{\citenamefont {Tsuji}\ \emph {et~al.}(2018)\citenamefont {Tsuji},
  \citenamefont {Otsuki},\ and\ \citenamefont {Katsuragi}}]{Tsuji2018}%
  \BibitemOpen
  \bibfield  {author} {\bibinfo {author} {\bibfnamefont {D.}~\bibnamefont
  {Tsuji}}, \bibinfo {author} {\bibfnamefont {M.}~\bibnamefont {Otsuki}}, \
  and\ \bibinfo {author} {\bibfnamefont {H.}~\bibnamefont {Katsuragi}},\
  }\href@noop {} {\bibfield  {journal} {\bibinfo  {journal} {Phys. Rev. Lett.}\
  }\textbf {\bibinfo {volume} {120}},\ \bibinfo {pages} {128001} (\bibinfo
  {year} {2018})}\BibitemShut {NoStop}%
\bibitem [{\citenamefont {Roering}\ \emph {et~al.}(1999)\citenamefont
  {Roering}, \citenamefont {Kirchner},\ and\ \citenamefont
  {Dietrich}}]{Roering1999}%
  \BibitemOpen
  \bibfield  {author} {\bibinfo {author} {\bibfnamefont {J.~J.}\ \bibnamefont
  {Roering}}, \bibinfo {author} {\bibfnamefont {J.~W.}\ \bibnamefont
  {Kirchner}}, \ and\ \bibinfo {author} {\bibfnamefont {W.~E.}\ \bibnamefont
  {Dietrich}},\ }\href@noop {} {\bibfield  {journal} {\bibinfo  {journal}
  {Water Resour. Res}\ }\textbf {\bibinfo {volume} {35}},\ \bibinfo {pages}
  {853} (\bibinfo {year} {1999})}\BibitemShut {NoStop}%
\bibitem [{\citenamefont {Pouliquen}(1999)}]{Pouliquen1999}%
  \BibitemOpen
  \bibfield  {author} {\bibinfo {author} {\bibfnamefont {O.}~\bibnamefont
  {Pouliquen}},\ }\href@noop {} {\bibfield  {journal} {\bibinfo  {journal}
  {Phys. Fluids}\ }\textbf {\bibinfo {volume} {11}},\ \bibinfo {pages} {542}
  (\bibinfo {year} {1999})}\BibitemShut {NoStop}%
\bibitem [{\citenamefont {Silbert}\ \emph {et~al.}(2001)\citenamefont
  {Silbert}, \citenamefont {Erta\ifmmode~\mbox{\c{s}}\else \c{s}\fi{}},
  \citenamefont {Grest}, \citenamefont {Halsey}, \citenamefont {Levine},\ and\
  \citenamefont {Plimpton}}]{Silbert2001}%
  \BibitemOpen
  \bibfield  {author} {\bibinfo {author} {\bibfnamefont {L.~E.}\ \bibnamefont
  {Silbert}}, \bibinfo {author} {\bibfnamefont {D.}~\bibnamefont
  {Erta\ifmmode~\mbox{\c{s}}\else \c{s}\fi{}}}, \bibinfo {author}
  {\bibfnamefont {G.~S.}\ \bibnamefont {Grest}}, \bibinfo {author}
  {\bibfnamefont {T.~C.}\ \bibnamefont {Halsey}}, \bibinfo {author}
  {\bibfnamefont {D.}~\bibnamefont {Levine}}, \ and\ \bibinfo {author}
  {\bibfnamefont {S.~J.}\ \bibnamefont {Plimpton}},\ }\href@noop {} {\bibfield
  {journal} {\bibinfo  {journal} {Phys. Rev. E}\ }\textbf {\bibinfo {volume}
  {64}},\ \bibinfo {pages} {051302} (\bibinfo {year} {2001})}\BibitemShut
  {NoStop}%
\bibitem [{\citenamefont {Andreotti}\ \emph {et~al.}(2013)\citenamefont
  {Andreotti}, \citenamefont {Forterre},\ and\ \citenamefont
  {Pouliquen}}]{Andreotti2013}%
  \BibitemOpen
  \bibfield  {author} {\bibinfo {author} {\bibfnamefont {B.}~\bibnamefont
  {Andreotti}}, \bibinfo {author} {\bibfnamefont {Y.}~\bibnamefont {Forterre}},
  \ and\ \bibinfo {author} {\bibfnamefont {O.}~\bibnamefont {Pouliquen}},\
  }\href@noop {} {\emph {\bibinfo {title} {Granular Media: Between Fluid and
  Solid}}}\ (\bibinfo  {publisher} {Cambridge University Press},\ \bibinfo
  {address} {Cambridge, U.K.},\ \bibinfo {year} {2013})\BibitemShut {NoStop}%
\bibitem [{\citenamefont {Gray}\ and\ \citenamefont
  {Edwards}(2014)}]{Gray2014}%
  \BibitemOpen
  \bibfield  {author} {\bibinfo {author} {\bibfnamefont {J.~M. N.~T.}\
  \bibnamefont {Gray}}\ and\ \bibinfo {author} {\bibfnamefont {A.~N.}\
  \bibnamefont {Edwards}},\ }\href@noop {} {\bibfield  {journal} {\bibinfo
  {journal} {J. Fluid Mech.}\ }\textbf {\bibinfo {volume} {755}},\ \bibinfo
  {pages} {503} (\bibinfo {year} {2014})}\BibitemShut {NoStop}%
\bibitem [{\citenamefont {Gaudel}\ \emph {et~al.}(2016)\citenamefont {Gaudel},
  \citenamefont {Kiesgen~de Richter}, \citenamefont {Louvet}, \citenamefont
  {Jenny},\ and\ \citenamefont {Skali-Lami}}]{Gaudel2016}%
  \BibitemOpen
  \bibfield  {author} {\bibinfo {author} {\bibfnamefont {N.}~\bibnamefont
  {Gaudel}}, \bibinfo {author} {\bibfnamefont {S.}~\bibnamefont {Kiesgen~de
  Richter}}, \bibinfo {author} {\bibfnamefont {N.}~\bibnamefont {Louvet}},
  \bibinfo {author} {\bibfnamefont {M.}~\bibnamefont {Jenny}}, \ and\ \bibinfo
  {author} {\bibfnamefont {S.}~\bibnamefont {Skali-Lami}},\ }\href@noop {}
  {\bibfield  {journal} {\bibinfo  {journal} {Phys. Rev. E}\ }\textbf {\bibinfo
  {volume} {94}},\ \bibinfo {pages} {032904} (\bibinfo {year}
  {2016})}\BibitemShut {NoStop}%
\bibitem [{\citenamefont {Tennakoon}\ and\ \citenamefont
  {Behringer}(1998)}]{Tennakoon1998}%
  \BibitemOpen
  \bibfield  {author} {\bibinfo {author} {\bibfnamefont {S.~G.~K.}\
  \bibnamefont {Tennakoon}}\ and\ \bibinfo {author} {\bibfnamefont {R.~P.}\
  \bibnamefont {Behringer}},\ }\href@noop {} {\bibfield  {journal} {\bibinfo
  {journal} {Phys. Rev. Lett.}\ }\textbf {\bibinfo {volume} {81}},\ \bibinfo
  {pages} {794} (\bibinfo {year} {1998})}\BibitemShut {NoStop}%
\bibitem [{\citenamefont {King}\ \emph {et~al.}(2000)\citenamefont {King},
  \citenamefont {Swift}, \citenamefont {Benedict},\ and\ \citenamefont
  {Routledge}}]{King2000}%
  \BibitemOpen
  \bibfield  {author} {\bibinfo {author} {\bibfnamefont {P.~J.}\ \bibnamefont
  {King}}, \bibinfo {author} {\bibfnamefont {M.~R.}\ \bibnamefont {Swift}},
  \bibinfo {author} {\bibfnamefont {K.~A.}\ \bibnamefont {Benedict}}, \ and\
  \bibinfo {author} {\bibfnamefont {A.}~\bibnamefont {Routledge}},\ }\href@noop
  {} {\bibfield  {journal} {\bibinfo  {journal} {Phys. Rev. E}\ }\textbf
  {\bibinfo {volume} {62}},\ \bibinfo {pages} {6982} (\bibinfo {year}
  {2000})}\BibitemShut {NoStop}%
\bibitem [{\citenamefont {Cundall}\ and\ \citenamefont
  {Strack}(1979)}]{Cundall1979}%
  \BibitemOpen
  \bibfield  {author} {\bibinfo {author} {\bibfnamefont {P.~A.}\ \bibnamefont
  {Cundall}}\ and\ \bibinfo {author} {\bibfnamefont {O.~D.~L.}\ \bibnamefont
  {Strack}},\ }\href@noop {} {\bibfield  {journal} {\bibinfo  {journal}
  {G$\acute{e}$otechnique}\ }\textbf {\bibinfo {volume} {29}},\ \bibinfo
  {pages} {47} (\bibinfo {year} {1979})}\BibitemShut {NoStop}%
\bibitem [{\citenamefont {Iikawa}\ \emph {et~al.}(2016)\citenamefont {Iikawa},
  \citenamefont {Bandi},\ and\ \citenamefont {Katsuragi}}]{Iikawa2016}%
  \BibitemOpen
  \bibfield  {author} {\bibinfo {author} {\bibfnamefont {N.}~\bibnamefont
  {Iikawa}}, \bibinfo {author} {\bibfnamefont {M.~M.}\ \bibnamefont {Bandi}}, \
  and\ \bibinfo {author} {\bibfnamefont {H.}~\bibnamefont {Katsuragi}},\
  }\href@noop {} {\bibfield  {journal} {\bibinfo  {journal} {Phys. Rev. Lett.}\
  }\textbf {\bibinfo {volume} {116}},\ \bibinfo {pages} {128001} (\bibinfo
  {year} {2016})}\BibitemShut {NoStop}%
\bibitem [{\citenamefont {Otsuki}\ and\ \citenamefont
  {Hayakawa}(2009)}]{Otsuki2009}%
  \BibitemOpen
  \bibfield  {author} {\bibinfo {author} {\bibfnamefont {M.}~\bibnamefont
  {Otsuki}}\ and\ \bibinfo {author} {\bibfnamefont {H.}~\bibnamefont
  {Hayakawa}},\ }\href@noop {} {\bibfield  {journal} {\bibinfo  {journal}
  {Phys. Rev. E}\ }\textbf {\bibinfo {volume} {80}},\ \bibinfo {pages} {011308}
  (\bibinfo {year} {2009})}\BibitemShut {NoStop}%
\bibitem [{\citenamefont {Zhou}\ \emph {et~al.}(2001)\citenamefont {Zhou},
  \citenamefont {Xu}, \citenamefont {Yu},\ and\ \citenamefont
  {Zulli}}]{Zhou2001}%
  \BibitemOpen
  \bibfield  {author} {\bibinfo {author} {\bibfnamefont {Y.~C.}\ \bibnamefont
  {Zhou}}, \bibinfo {author} {\bibfnamefont {B.~H.}\ \bibnamefont {Xu}},
  \bibinfo {author} {\bibfnamefont {A.~B.}\ \bibnamefont {Yu}}, \ and\ \bibinfo
  {author} {\bibfnamefont {P.}~\bibnamefont {Zulli}},\ }\href@noop {}
  {\bibfield  {journal} {\bibinfo  {journal} {Phys. Rev. E}\ }\textbf {\bibinfo
  {volume} {64}},\ \bibinfo {pages} {021301} (\bibinfo {year}
  {2001})}\BibitemShut {NoStop}%
\bibitem [{\citenamefont {Zhou}\ \emph {et~al.}(2002)\citenamefont {Zhou},
  \citenamefont {Xu}, \citenamefont {Yu},\ and\ \citenamefont
  {Zulli}}]{Zhou2002}%
  \BibitemOpen
  \bibfield  {author} {\bibinfo {author} {\bibfnamefont {Y.}~\bibnamefont
  {Zhou}}, \bibinfo {author} {\bibfnamefont {B.}~\bibnamefont {Xu}}, \bibinfo
  {author} {\bibfnamefont {A.}~\bibnamefont {Yu}}, \ and\ \bibinfo {author}
  {\bibfnamefont {P.}~\bibnamefont {Zulli}},\ }\href@noop {} {\bibfield
  {journal} {\bibinfo  {journal} {Powder Technology}\ }\textbf {\bibinfo
  {volume} {125}},\ \bibinfo {pages} {45 } (\bibinfo {year}
  {2002})}\BibitemShut {NoStop}%
\bibitem [{\citenamefont {Lueptow}\ \emph {et~al.}(2000)\citenamefont
  {Lueptow}, \citenamefont {Akonur},\ and\ \citenamefont
  {Shinbrot}}]{Lueptow2000}%
  \BibitemOpen
  \bibfield  {author} {\bibinfo {author} {\bibfnamefont {R.~M.}\ \bibnamefont
  {Lueptow}}, \bibinfo {author} {\bibfnamefont {A.}~\bibnamefont {Akonur}}, \
  and\ \bibinfo {author} {\bibfnamefont {T.}~\bibnamefont {Shinbrot}},\
  }\href@noop {} {\bibfield  {journal} {\bibinfo  {journal} {Experiments in
  Fluids}\ }\textbf {\bibinfo {volume} {28}},\ \bibinfo {pages} {183} (\bibinfo
  {year} {2000})}\BibitemShut {NoStop}%
\bibitem [{\citenamefont {Ghazavi}\ and\ \citenamefont
  {Mollanouri}(2008)}]{Ghazavi2008}%
  \BibitemOpen
  \bibfield  {author} {\bibinfo {author} {\bibfnamefont {M.}~\bibnamefont
  {Ghazavi}, \bibfnamefont {M.and~Hosseini}}\ and\ \bibinfo {author}
  {\bibfnamefont {M.}~\bibnamefont {Mollanouri}},\ }\href@noop {} {\bibfield
  {journal} {\bibinfo  {journal} {The 12th International Conference of
  International Association for Computer Methods and Advances in Geomechanics}\
  ,\ \bibinfo {pages} {1272 }} (\bibinfo {year} {2008})}\BibitemShut {NoStop}%
\bibitem [{\citenamefont {da~Cruz}\ \emph {et~al.}(2005)\citenamefont
  {da~Cruz}, \citenamefont {Emam}, \citenamefont {Prochnow}, \citenamefont
  {Roux},\ and\ \citenamefont {Chevoir}}]{Cruz2005}%
  \BibitemOpen
  \bibfield  {author} {\bibinfo {author} {\bibfnamefont {F.}~\bibnamefont
  {da~Cruz}}, \bibinfo {author} {\bibfnamefont {S.}~\bibnamefont {Emam}},
  \bibinfo {author} {\bibfnamefont {M.}~\bibnamefont {Prochnow}}, \bibinfo
  {author} {\bibfnamefont {J.-N.}\ \bibnamefont {Roux}}, \ and\ \bibinfo
  {author} {\bibfnamefont {F.}~\bibnamefont {Chevoir}},\ }\href@noop {}
  {\bibfield  {journal} {\bibinfo  {journal} {Phys. Rev. E}\ }\textbf {\bibinfo
  {volume} {72}},\ \bibinfo {pages} {021309} (\bibinfo {year}
  {2005})}\BibitemShut {NoStop}%
\bibitem [{\citenamefont {Roering}(2004)}]{Roering2004}%
  \BibitemOpen
  \bibfield  {author} {\bibinfo {author} {\bibfnamefont {J.~J.}\ \bibnamefont
  {Roering}},\ }\href@noop {} {\bibfield  {journal} {\bibinfo  {journal} {Earth
  Surf. Process. Landforms}\ }\textbf {\bibinfo {volume} {29}},\ \bibinfo
  {pages} {1597} (\bibinfo {year} {2004})}\BibitemShut {NoStop}%
\bibitem [{\citenamefont {Gabet}(2003)}]{Gabet2003}%
  \BibitemOpen
  \bibfield  {author} {\bibinfo {author} {\bibfnamefont {E.~J.}\ \bibnamefont
  {Gabet}},\ }\href@noop {} {\bibfield  {journal} {\bibinfo  {journal} {Journal
  of Geophysical Research: Solid Earth}\ }\textbf {\bibinfo {volume} {108}},\
  \bibinfo {pages} {2049} (\bibinfo {year} {2003})}\BibitemShut {NoStop}%
\bibitem [{\citenamefont {Roering}\ \emph {et~al.}(2001)\citenamefont
  {Roering}, \citenamefont {Kirchner}, \citenamefont {Sklar},\ and\
  \citenamefont {Dietrich}}]{Roering2001}%
  \BibitemOpen
  \bibfield  {author} {\bibinfo {author} {\bibfnamefont {J.~J.}\ \bibnamefont
  {Roering}}, \bibinfo {author} {\bibfnamefont {J.~W.}\ \bibnamefont
  {Kirchner}}, \bibinfo {author} {\bibfnamefont {L.~S.}\ \bibnamefont {Sklar}},
  \ and\ \bibinfo {author} {\bibfnamefont {W.~E.}\ \bibnamefont {Dietrich}},\
  }\href@noop {} {\bibfield  {journal} {\bibinfo  {journal} {Geology}\ }\textbf
  {\bibinfo {volume} {29}},\ \bibinfo {pages} {143} (\bibinfo {year}
  {2001})}\BibitemShut {NoStop}%
\bibitem [{\citenamefont {Furbish}\ \emph {et~al.}(2008)\citenamefont
  {Furbish}, \citenamefont {Schmeeckle},\ and\ \citenamefont
  {Roering}}]{Furbish2008}%
  \BibitemOpen
  \bibfield  {author} {\bibinfo {author} {\bibfnamefont {D.~J.}\ \bibnamefont
  {Furbish}}, \bibinfo {author} {\bibfnamefont {M.~W.}\ \bibnamefont
  {Schmeeckle}}, \ and\ \bibinfo {author} {\bibfnamefont {J.~J.}\ \bibnamefont
  {Roering}},\ }\href@noop {} {\bibfield  {journal} {\bibinfo  {journal} {Earth
  Surf. Process. Landforms}\ }\textbf {\bibinfo {volume} {33}},\ \bibinfo
  {pages} {2108} (\bibinfo {year} {2008})}\BibitemShut {NoStop}%
\bibitem [{\citenamefont {BenDror}\ and\ \citenamefont
  {Goren}(2018)}]{BenDror2018}%
  \BibitemOpen
  \bibfield  {author} {\bibinfo {author} {\bibfnamefont {E.}~\bibnamefont
  {BenDror}}\ and\ \bibinfo {author} {\bibfnamefont {L.}~\bibnamefont
  {Goren}},\ }\href@noop {} {\bibfield  {journal} {\bibinfo  {journal} {Journal
  of Geophysical Research: Earth Surface}\ }\textbf {\bibinfo {volume} {123}},\
  \bibinfo {pages} {924} (\bibinfo {year} {2018})}\BibitemShut {NoStop}%
\bibitem [{\citenamefont {Ferdowsi}\ \emph {et~al.}(2018)\citenamefont
  {Ferdowsi}, \citenamefont {Ortiz},\ and\ \citenamefont
  {Jerolmack}}]{Ferdowsi2018}%
  \BibitemOpen
  \bibfield  {author} {\bibinfo {author} {\bibfnamefont {B.}~\bibnamefont
  {Ferdowsi}}, \bibinfo {author} {\bibfnamefont {C.~P.}\ \bibnamefont {Ortiz}},
  \ and\ \bibinfo {author} {\bibfnamefont {D.~J.}\ \bibnamefont {Jerolmack}},\
  }\href@noop {} {\bibfield  {journal} {\bibinfo  {journal} {PNAS}\ }\textbf
  {\bibinfo {volume} {115}},\ \bibinfo {pages} {4827} (\bibinfo {year}
  {2018})}\BibitemShut {NoStop}%
\bibitem [{\citenamefont {Jaeger}\ \emph {et~al.}(1996)\citenamefont {Jaeger},
  \citenamefont {Nagel},\ and\ \citenamefont {Behringer}}]{Jaeger1996}%
  \BibitemOpen
  \bibfield  {author} {\bibinfo {author} {\bibfnamefont {H.~M.}\ \bibnamefont
  {Jaeger}}, \bibinfo {author} {\bibfnamefont {S.~R.}\ \bibnamefont {Nagel}}, \
  and\ \bibinfo {author} {\bibfnamefont {R.~P.}\ \bibnamefont {Behringer}},\
  }\href@noop {} {\bibfield  {journal} {\bibinfo  {journal} {Rev. Mod. Phys.}\
  }\textbf {\bibinfo {volume} {68}},\ \bibinfo {pages} {1259} (\bibinfo {year}
  {1996})}\BibitemShut {NoStop}%
\bibitem [{\citenamefont {Richman}(1993)}]{Richman1993}%
  \BibitemOpen
  \bibfield  {author} {\bibinfo {author} {\bibfnamefont {M.}~\bibnamefont
  {Richman}},\ }\href@noop {} {\bibfield  {journal} {\bibinfo  {journal}
  {Mechanics of Materials}\ }\textbf {\bibinfo {volume} {16}},\ \bibinfo
  {pages} {211 } (\bibinfo {year} {1993})}\BibitemShut {NoStop}%
\bibitem [{\citenamefont {Jenkins}\ and\ \citenamefont
  {Berzi}(2012)}]{Jenkins2012}%
  \BibitemOpen
  \bibfield  {author} {\bibinfo {author} {\bibfnamefont {J.~T.}\ \bibnamefont
  {Jenkins}}\ and\ \bibinfo {author} {\bibfnamefont {D.}~\bibnamefont
  {Berzi}},\ }\href@noop {} {\bibfield  {journal} {\bibinfo  {journal}
  {Granular Matter}\ }\textbf {\bibinfo {volume} {14}},\ \bibinfo {pages} {79}
  (\bibinfo {year} {2012})}\BibitemShut {NoStop}%
\bibitem [{\citenamefont {Pak}\ and\ \citenamefont
  {Behringer}(1993)}]{Pak1993}%
  \BibitemOpen
  \bibfield  {author} {\bibinfo {author} {\bibfnamefont {H.~K.}\ \bibnamefont
  {Pak}}\ and\ \bibinfo {author} {\bibfnamefont {R.~P.}\ \bibnamefont
  {Behringer}},\ }\href@noop {} {\bibfield  {journal} {\bibinfo  {journal}
  {Phys. Rev. Lett.}\ }\textbf {\bibinfo {volume} {71}},\ \bibinfo {pages}
  {1832} (\bibinfo {year} {1993})}\BibitemShut {NoStop}%
\end{thebibliography}%

\end{document}